\documentclass[reprint,amsmath,amssymb,aps,pra,superscriptaddress]{revtex4-2}

\usepackage{graphicx}
\usepackage{dcolumn}
\usepackage{bm}
\usepackage{amsmath}

\usepackage{xcolor}

\begin{document}


\title{Imaginary couplings in non-Hermitian coupled-mode theory: \\ Effects on exceptional points of optical resonators}



\author{Kenta Takata}
\email{kenta.takata.ke@hco.ntt.co.jp}
\affiliation{Nanophotonics Center, NTT Corporation, Atsugi, Kanagawa, Japan}
\affiliation{NTT Basic Research Laboratories, NTT Corporation, Atsugi, Kanagawa, Japan}

\author{Nathan Roberts}
\altaffiliation[Currently at ]{University of Bath, Claverton Down, Bath, UK.}
\affiliation{NTT Basic Research Laboratories, NTT Corporation, Atsugi, Kanagawa, Japan}

\author{Akihiko Shinya}
\affiliation{Nanophotonics Center, NTT Corporation, Atsugi, Kanagawa, Japan}
\affiliation{NTT Basic Research Laboratories, NTT Corporation, Atsugi, Kanagawa, Japan}

\author{Masaya Notomi}
\affiliation{Nanophotonics Center, NTT Corporation, Atsugi, Kanagawa, Japan}
\affiliation{NTT Basic Research Laboratories, NTT Corporation, Atsugi, Kanagawa, Japan}
\affiliation{Department of Physics, Tokyo Institute of Technology, Meguro-ku, Tokyo, Japan}


\date{\today}

\begin{abstract}
Exceptional point (EP) degeneracies in coupled cavities with gain and loss provide on-chip photonic devices with unconventional features and performance. However, such systems with realistic structures often miss the exact EPs even in simulation, and the mechanism of this EP disruption has yet to be thoroughly identified. Here, we extend the coupled-mode theory of one-dimensional non-Hermitian resonator arrays to study the effects of the imaginary part of the inter-cavity coupling, which is a second-order term and attributed to material amplification, absorption, and radiation. By taking an appropriate gauge for the model, we clarify that the imaginary coupling components have a symmetric form in the effective Hamiltonian and hence represent non-Hermiticity. These additional factors can lift the gain- and loss-based EP degeneracies. However, they are proportional to the sum of the imaginary permittivities for involved cavity pairs. Thus, when the amplification and absorption of adjacent cavities are balanced, their contribution to the imaginary coupling is canceled, and the EP singularity can be restored. Radiation-induced imaginary couplings measure the change in net radiation loss by the interference between cavity modes. Their impact on the EP can also be counteracted by small cavity resonance detuning even in loss-biased cases. We show and analyze eligible simulation examples based on photonic crystal nanocavities, and highlight design of an ideal EP degeneracy that is protected by generalized PT symmetry and induced by radiation.
\end{abstract}


\maketitle

\section{\label{sec:introduction} Introduction}
Exceptional points (EPs) in parity-time-symmetric (PT-symmetric) optical systems are the degeneracies of eigenmodes originating from the contrast of their imaginary parts of refractive indices, namely distributed gain and loss \cite{Kato1995,Bender1998,Bender2002,Heiss2012,Berry2004,Makris2008,Musslimani2008,Klaiman2008,Feng2017non,ElGanainy2018non,Ozdemir2019par,Miri2019}. Ideal EPs make not only some spectral eigenvalues but also corresponding eigenstates identical. This EP degeneracy induces directional responses \cite{Lin2011,Regensburger2012,Feng2013,Gao2017}, single chirality \cite{Miao2016,Peng2016,Zhang2020}, and enhancement of photonic local density of states \cite{Lin2016,Pick2017gen,Pick2017enh,Takata2021}. In addition, EPs correspond to branch points of the spectral eigenvalues that behave as complex radical multifunctions. Thus, the system around them undergoes a peculiar phase transition from extended to localized modes with singularity \cite{Guo2009,Ruter2010}. This EP transition enables optical isolation \cite{Peng2014,Chang2014}, modal control of lasers \cite{Feng2014sin,Hodaei2014,Brandstetter2014,Peng2014los,Wong2016}, and enhanced sensitivity \cite{Hodaei2017enh,Chen2017exc}, to name a few. Moreover, encircling an EP can give rise to eigenmode switching \cite{Dembowski2001,Zhong2018} and asymmetric mode conversion \cite{Doppler2016,Hassan2017}. EPs also exhibit unconventional charge vortices, topological states, and symmetry protection \cite{Leykam2017,Zhou2018,Takata2018,Malzard2015,Pan2018,Okugawa2019,Yoshida2019,Kawabata2019,Ota2020,Parto2021}. 

Despite their various intriguing properties, photonic EPs are single points in continuous parameter spaces and can hence be lifted by small perturbation, which is common in practical systems. The resultant unavailability of the exact EP hampers not only the unconventional responses based on the degenerate eigenstate but also the divergent change in frequency required for sensing applications. In coupled cavities (waveguides), this problem is usually attributed to the undesired detuning of their resonance frequencies (propagation constants), which stems from structural disorder and biased distributions of heat and active carriers. On the other hand, the exact EPs often dissolve even in simulations that do not take these effects into consideration \cite{Takata2017,Yao2019}. Thus, the entire mechanism of their disappearance should be explored.

Gain- and loss-induced EPs in evanescently coupled photonic lattices are mostly analyzed with the system's effective Hamiltonian in the coupled-mode theory (CMT) within the first order of the coupling coefficient \cite{Yariv1999,Xu2000,Poon2007}. This framework seems equivalent to a variational approach \cite{Wu2016} and includes the on-site gain and loss, mode detuning, and real coupling terms. However, it does not cover the lifting of such EPs in systems with no detuning.

In fact, it was pointed out phenomenologically that general complex couplings might be responsible for this EP disruption \cite{Benisty2012}. An elaborate CMT for uniformly lossy waveguide arrays \cite{Golshani2014} also revealed the existence of a small imaginary component of the waveguide coupling that had been missed in the literature. Moreover, a perturbation analysis for two coupled planar waveguides \cite{Nguyen2016} indicated asymmetric complex couplings for guided-mode basis, which actually resulted in smoothed EP transitions of eigenvalues. However, it is still unclear how the material gain, absorption loss, and radiation of on-chip coupled resonators (waveguides) exactly contribute to the system's coupled-mode equations (CMEs). A comprehensive picture of such non-Hermitian CMEs is needed to understand how the EP degeneracies are actually lost and can possibly be restored. Dependable design principles for realistic devices with exact EPs are also desired.

In this study, we extend the non-Hermitian coupled-mode formalism for periodic units of two dielectric resonators with gain and loss. We clarify that the amplification, absorption, and radiation of cavity fields all give rise to finite imaginary parts of the inter-cavity couplings. These \textit{imaginary coupling} terms take a symmetric form in the effective Hamiltonian and hence represents non-Hermiticity, as it works as the source of anti-PT-symmetric EPs in other dissipative systems \cite{Chen2017exc,Fan2020}. It generally lifts the EPs based on on-site gain and/or loss. We also show, however, that if the unit cell has balanced imaginary refractive indices in the cavities, their contribution to the imaginary coupling is canceled. We simulate complex band structures of photonic crystal cavities based on buried heterostructures with amplification and absorption, and confirm that our imaginary coupling explains well the disappearance and revival of their EP depending on the parameters.

Coupled cavities with radiation loss are found to be formally described by the same CMEs as those with material gain and loss. Here, we reveal that the radiation-induced imaginary couplings denote the change in net loss of coupled modes by the interference of radiation fields. They may have complicated non-local features unlike the permittivity-induced effects, because radiation fields inherit phase coherence of coupled modes and are free from exponential decay in air. We examine a two-cavity system where the model is exact and show a modified EP condition that additionally requires the balance between the imaginary coupling and cavity frequency detuning. Such EPs respect a generalized PT symmetry and exhibit topological robustness to continuous changes in parameters. We also simulate two coupled point-defect photonic crystal cavities and identify their ideal EP induced by radiation. Structural modulation of the cavities results in both the contrast of their solitary radiation loss and variation in their complex inter-cavity coupling. Although the latter continuously dislocates the EP in the CMT parameter space, we successfully find a condition where coexisting small resonance detuning compensates for the imaginary coupling and thus the system reaches a singular EP degeneracy.

\section{\label{sec:overview} Overview and article structure}
\begin{figure}[t!]
	\includegraphics[width=8.0cm]{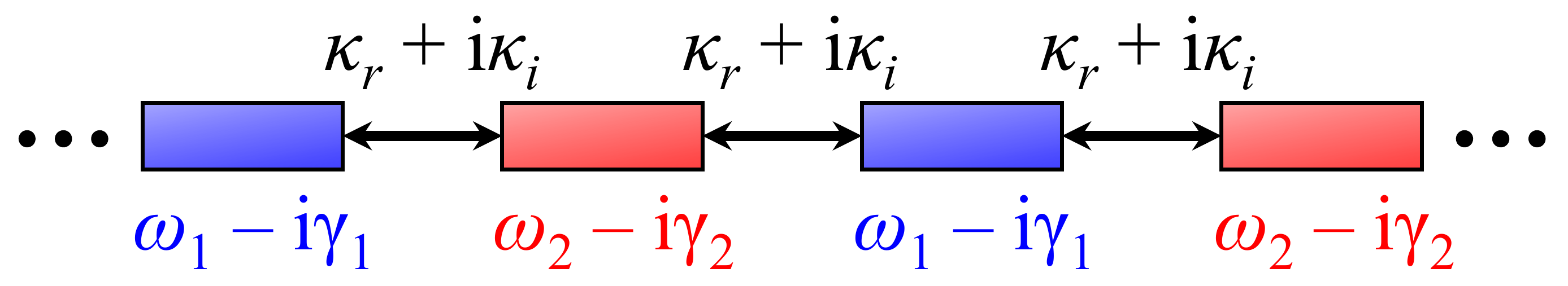}
	\caption{\label{fig:concept} Conceptual schematic of the system. It comprises pairs of equally spaced single-mode cavities with resonance frequencies $(\omega_1, \omega_2)$ and on-site gain and loss $(\gamma_1, \gamma_2)$. In our CMT, the gain and loss induce a small direction-independent imaginary coupling $i \kappa_{i} \in i \mathbb{R}$, in addition to the Hermitian evanescent coupling $\kappa_{r} \in \mathbb{R}$.}
\end{figure}

We consider one-dimensional periodic systems of coupled single-mode optical resonators (Fig. \ref{fig:concept}). Their unit cells have two generally distinct cavities with unperturbed resonance frequencies $(\omega_1, \omega_2)$ and on-site modal gain and/or loss $(\gamma_1, \gamma_2)$. All the cavities are supposed to be equally spaced so that they will be uniformly coupled by evanescent waves.

In this study, we show that the CMEs for such systems can be written as
\begin{align}
	&-i \frac{\textrm{d}a_{2 h - 1}}{\textrm{d}t} = \left( \delta - i \gamma_1 \right) a_{2 h - 1} - \left( \kappa_{r} + i \kappa_{i} \right)\left(a_{2 h - 2} + a_{2 h} \right), \nonumber \\
	&-i \frac{\textrm{d}a_{2 h}}{\textrm{d}t} = \left( -\delta - i \gamma_2 \right) a_{2 h} - \left( \kappa_{r} + i \kappa_{i} \right)\left( a_{2 h - 1} + a_{2 h + 1} \right), \label{eq:CME}
\end{align}
where $a_n$ is the mode amplitude for cavity $n$, and $h = 1, 2 \ldots$ the index of the unit cells. The real coupling term $\kappa_{r} \in \mathbb{R}$ measures the lossless energy exchange between cavities and is common in the literature. In addition, we derive an imaginary counterpart $i \kappa_{i} \in i \mathbb{R}$, which denotes the interplay between the photonic hopping and material gain and loss, or the modulation of net radiation loss by the interference of radiation fields. The evanescently coupled cavities indicate that $\kappa_r \gg \kappa_i$. $\pm \delta = \pm (\omega_1 - \omega_2)/2$ are the relative mode frequencies of the cavities to their average $\omega_0 = (\omega_1 + \omega_2)/2$. This detuning parameter might be controlled to reach the EP even under the existence of $\kappa_i$.
In Sec. \ref{sec:CMT}, we derive the CMEs for coupled dielectric resonators with finite imaginary parts of the permittivities of the cavity media in order to investigate how the imaginary coupling terms appear and depend on the material gain and loss. The system under the periodic boundary condition exhibits the following complex eigenfrequency detuning:
\begin{align}
	\Delta \omega (k) = &- i \frac{\gamma_1 + \gamma_2}{2} \nonumber \\
	&\pm \sqrt{4 \left(\kappa_{r} + i \kappa_{i}\right)^2 \cos^2 \frac{k L}{2} - \bigg(\frac{\gamma_1 - \gamma_2}{2} + i \delta \bigg)^2},
	\label{eq:bulkeigenvalue0}
\end{align}
where $k$ is the Bloch wave number and $L$ is the lattice constant of the unit cells. When $\kappa_i = \delta = 0$, the system respects PT symmetry and has exact EPs for $|\gamma_1 - \gamma_2| \le 4|\kappa_r|$. However, finite $\kappa_i$ generally washes out such degeneracies from the band structure, even though the constituent cavity resonances are coincident, i.e., $\delta = 0$.

In Sec. \ref{sec:BHsystem}, we show the power of our CMT formalism with a simulation of buried-heterostructure photonic crystal cavity arrays. The simulation result clarifies that the amplification- and absorption-based imaginary couplings are proportional to the sum of the imaginary permittivities of the involved cavities, being consistent with our analytic derivation. The balanced material gain and loss in the unit cell hence cancel their contribution to the imaginary coupling and restore the EP, as long as radiation effects and $\delta$ are negligible.

In Sec. \ref{sec:CMTRad}, we provide a CMT for passive coupled cavities with contrast of their radiation loss. By modeling the radiation effects with a virtual absorber placed at infinity, we obtain CMEs with the same form as Eq. (\ref{eq:CME}). Here, we focus on a system of two cavities, because some non-local imaginary couplings, implied in Sec. \ref{sec:BHsystem}, may be present in radiation-based systems with more cavities. For clarity, we write the CMEs here as
\begin{eqnarray}
	-i \frac{\textrm{d}a_1}{\textrm{d}t} &=& \left( \delta - i \Gamma_1 \right) a_1 - \left( \kappa_{r} + i K_{i} \right) a_2, \nonumber \\ 
	-i \frac{\textrm{d}a_2}{\textrm{d}t} &=& \left( -\delta - i \Gamma_2 \right) a_2 - \left( \kappa_{r} + i K_{i} \right)a_1, \label{eq:CMErad}
\end{eqnarray}
where ($\Gamma_1, \Gamma_2)$ and $K_i$ denote the cavities' radiation loss and radiation-induced imaginary coupling. Its theoretical eigenfrequency detuning $\Delta \omega_{\pm}$ reads
\begin{equation}
	\Delta \omega_{\pm} = - i\frac{\Gamma_1 + \Gamma_2}{2}
	\pm \sqrt{ \left(\kappa_{r} + i K_{i}\right)^2 - \bigg(\frac{\Gamma_1 - \Gamma_2}{2} + i \delta \bigg)^2 }. \label{eq:twocaveigenvalue0}
\end{equation}
Again, finite $K_i$ or $\delta$ generally lifts the EPs of $\Delta \omega_{\pm}$ in the non-Hermitian phase transition where the loss contrast $l \equiv (\Gamma_1 - \Gamma_2)/2$ is varied. However, we can identify such displaced EPs in the two-parameter space with $\delta$ and $l$. We also discuss their topological robustness based on generalized PT symmetry \cite{Kawabata2019}.

In Sec. \ref{sec:RBEP}, we theoretically demonstrate the radiation-based EP in two coupled Si photonic crystal nanocavities. By adjusting two distinct structural parameters, we control both $\delta$ and $l$ and find a parameter trajectory where the system respects the general PT symmetry. As a result, we reach an ideal EP accompanied with the singular coalescence of the eigenvalues $\Delta \omega_{\pm}$ and balanced $K_i$ and $\delta$. Our CMT explains the entire simulation result without any notable discrepancy that would suggest unexpected factors.

We discuss the applicability of our CMT, and conclude our study in Sec. \ref{sec:conclusion}.

\section{\label{sec:CMT} Non-Hermitian coupled-mode theory: effects of material amplification and absorption} 
\subsection{Derivation of coupled-mode equations}
\begin{figure}[t!]
	\includegraphics[width=8.6cm]{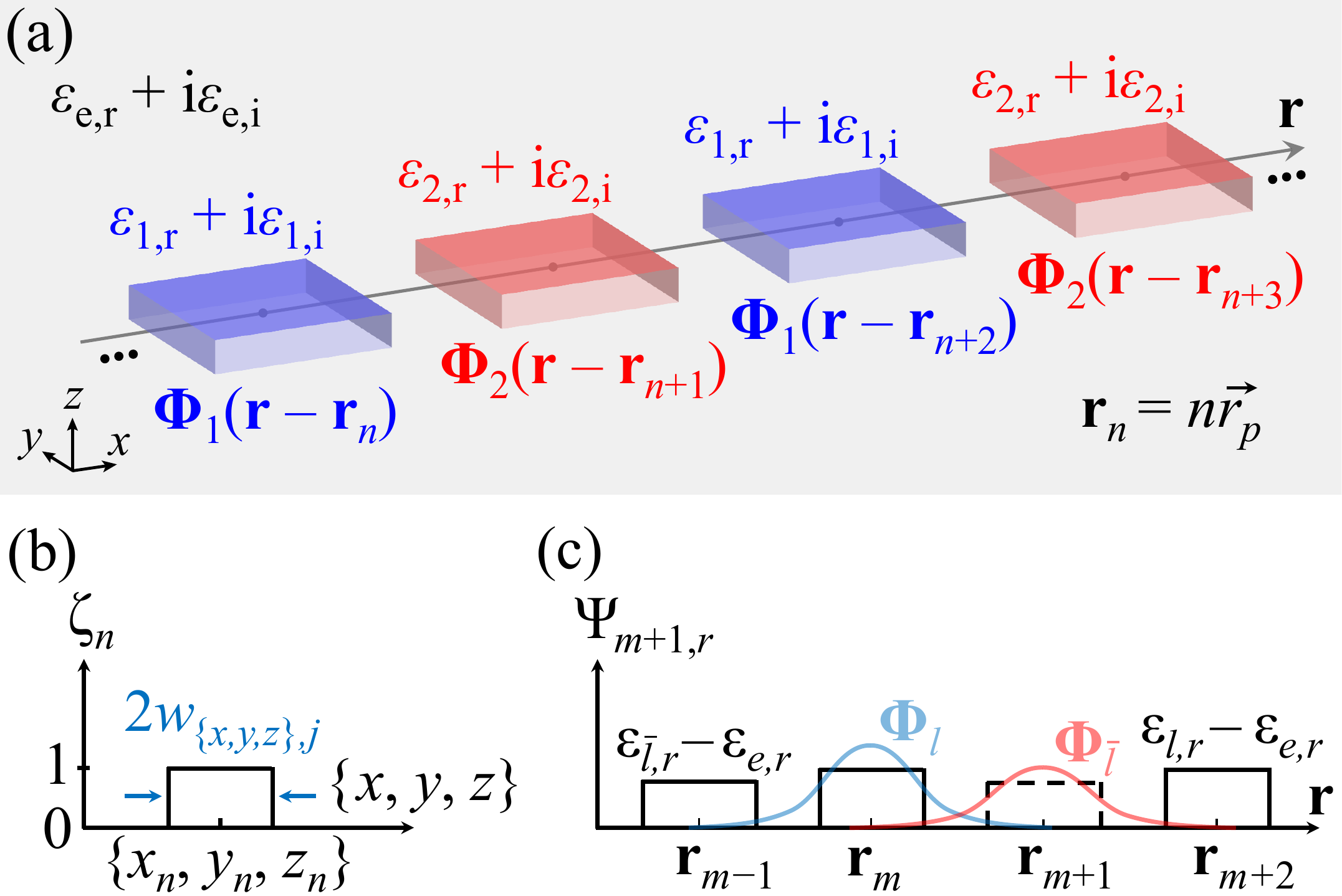}
	\caption{\label{fig:system} (a) Non-Hermitian coupled dielectric resonators. The unit cell comprises two equally aligned cavities with complex dielectric constants of $\{\epsilon_{j,r}+i\epsilon_{j,i} \}$ and unperturbed cavity modes $\{\bm{\Phi}_j\}$, where $j = 1, 2$ is the cavity modular index. The system is buried in a boundless cladding material with a permittivity of $\epsilon_{e,r} + i \epsilon_{e,i}$. (b) The function $\zeta_{n}({\bf r})$ in Eq. (\ref{eq:zeta}) that specifies the domain of cavity $n$. It is centered at ${\bf r}_n = (x_n, y_n, z_n)$ and has widths of $\{2w_{x, j}, 2w_{y, j}, 2w_{z, j}\}$ in $\{x, y, z\}$ directions. (c) The relative real permittivity distribution $\Psi_{m + 1, r} ({\bf r})$ based on Eq. (\ref{eq:Psi}) and schematic of the real coupling $\kappa_{m + 1, m}$ [Eq. (\ref{eq:realcoupling1})] from $\bm{\Phi}_{\bar{l}}({\bf r} - {\bf r}_{m+1})$ to $\bm{\Phi}_{l}({\bf r} - {\bf r}_{m})$, with $l$ and $\bar{l}$ being the modular indices for cavity $m$ and $m + 1$. $\Psi_{m + 1, r}$ lacks the contribution of the $m+1$th cavity (dashed lines) and hence extracts the inner product $\bm{\Phi}_{l} \cdot \bm{\Phi}_{\bar{l}}$ just within the $m$th one.}
\end{figure}

Here, we rigorously derive our CMEs [Eq. (\ref{eq:CME})] for active dielectric resonators and clarify how the permittivity-based imaginary couplings appear. The considered system is shown schematically in Fig. \ref{fig:system}(a).
It comprises unit cells of two rectangular cavities defined by heterostructures with high real specific permittivities $\{\epsilon _{j,r}\}$ compared to that of the boundless cladding material $\epsilon_{e,r}$. Here, the modular cavity index $j$ is 
\begin{equation}
	j =
	\begin{cases}
		1 & (n \ \textrm{mod} \ 2 = 1),\\
		2 & (n \ \textrm{mod} \ 2 = 0).\label{eq:definition_j}
	\end{cases}
\end{equation}
The two constituent cavities in three-dimensional Cartesian coordinates can have different sizes $2w_{x, j}$, $2w_{y, j}$ and $2w_{z, j}$ in $x$, $y$, and $z$ directions, respectively. The cavity media also have generally distinct imaginary parts of permittivity $\{\epsilon _{j,i}\}$. We postulate that the structural and material discrepancy among the cavities is sufficiently small so that the considered cavity modes have the same order and symmetric property. The position of the $n$th resonator is denoted as ${\bf r}_n \equiv (x_n, y_n, z_n) = n\vec{r_p}$, where the displacement between any adjacent cavities is equal and defined as $\vec{r_p}$. The relative permittivity distribution of the system with $N$ cavities can hence be written as
\begin{align}
	\epsilon ({\bf r}) =& \epsilon_{e,r} + i \epsilon_{e,i} \nonumber \\*
	& + \sum_{n=1}^{N} \left[(\epsilon_{n,r} - \epsilon_{e,r}) + i (\epsilon_{n,i} - \epsilon_{e,i}) \right] \zeta_{n}({\bf r}),  \label{eq:permittivity} \\
	\zeta_{n}({\bf r}) = & [H(\Delta x_n + w_{x,j}) - H(\Delta x_n - w_{x,j})] \nonumber \\*
	& \times [H(\Delta y_n + w_{y,j}) - H(\Delta y_n - w_{y,j})] \nonumber \\*
	& \times [H(\Delta z_n + w_{z,j}) - H(\Delta z_n - w_{z,j})], \label{eq:zeta}
\end{align}
with $\epsilon_{n,r} = \epsilon_{j,r} \in \mathbb{R}$ and $\epsilon_{n,i} = \epsilon_{j,i} \in \mathbb{R}$ according to Eq. (\ref{eq:definition_j}), $\Delta x_n \equiv x - x_n$, $\Delta y_n \equiv y - y_n$, and $\Delta z_n \equiv z - z_n$. $\zeta_{n}({\bf r})$ specifies the $n$th cavity with the products of Heaviside functions $H(\cdot)$ \cite{Golshani2014}; see also Fig. \ref{fig:system}(b).

Our CMT uses the electric eigenmodes ${\bf E}_{j}({\bf r}, t)$ for the constitutive single cavities ($j = 1, 2$), each of which is assumed here to be completely transparent ($\epsilon_{j,i} = \epsilon_{e,i} = 0$) and isolated in the cladding material. The time-domain Maxwell equation for electric fields ${\bf E}({\bf r}, t)$ is
\begin{equation}
	\nabla \times \nabla \times {\bf E}({\bf r}, t) + \frac{1}{c^2} \epsilon ({\bf r}) \frac{\partial^2 {\bf E}({\bf r}, t)}{\partial t^2} = 0, \label{eq:temporalMaxwell}
\end{equation}
where $c$ is the speed of light in a vacuum. Here, ${\bf E}_{j}({\bf r}, t)$ can be decomposed into a dynamical factor $\exp (i \omega_j t)$ and the vector-field components $\bm{\Phi}_j({\bf r})$ that satisfy the spatial Maxwell equation derived from Eq. (\ref{eq:temporalMaxwell}), namely
\begin{eqnarray}
	{\bf E}_{j}({\bf r}, t) = \bm{\Phi}_j({\bf r}) \exp (i \omega_j t), \label{eq:cavitymode} \\
	\nabla \times \nabla \times \bm{\Phi}_j({\bf r}) = \frac{\omega_j^2}{c^2} \epsilon_{j}({\bf r})  \bm{\Phi}_j({\bf r}), \label{eq:cavityspatialMaxwell} \\
	\epsilon_{j}({\bf r}) = \epsilon_{e,r} + (\epsilon_{j,r} - \epsilon_{e,r}) \zeta_{0, j}({\bf r}), \label{eq:cavitypermittivity}
\end{eqnarray}
\begin{align}
	\zeta_{0, j}({\bf r}) = & [H(x + w_{x,j}) - H(x - w_{x,j})] \nonumber \\*
	& \times [H(y + w_{y,j}) - H(y - w_{y,j})] \nonumber \\*
	& \times [H(z + w_{z,j}) - H(z - w_{z,j})], \label{eq:cavityzeta}
\end{align}
where $\epsilon_{j}({\bf r})$ is the permittivity distribution of cavity medium $j$ singly placed at the origin, and $\omega_j$ is the modal angular eigenfrequency for cavity $j$. Here, the complex refractive index is written as $n_{j, r} - i n_{j, i}$ in this notation.

The key of our formulation is to determine the gauge of the basis functions $\{{\bf E}_{j}({\bf r}, t)\}$. They are lossless and thus $\omega_j \in \mathbb{R}$, because of the condition that each of the solitary cavity media is transparent and its cladding layer is infinitely thick [Eqs. (\ref{eq:cavitypermittivity}) and (\ref{eq:cavityzeta})]. This means that  both Eq. (\ref{eq:cavityspatialMaxwell}) and its solution can be real, i.e., $\forall {\bf r}: \ \bm{\Phi}_j({\bf r}) \in \mathbb{R}^3$. We explicitly take this gauge so that the overlap integrals between any basis cavity modes also become real in determining the form of the CMEs.

In the coupled-mode approximation, the fields of the entire non-Hermitian resonator array are expanded with the displaced lossless cavity modes. We hence write ${\bf E}({\bf r}, t)$ as
\begin{equation}
	{\bf E}({\bf r}, t) = \left[ \sum_{n=1}^{N} a_n(t) \bm{\Phi}_j({\bf r} - {\bf r}_n) \right] \exp (i \omega_0 t). \label{eq:CMapprox}
\end{equation}
Here, $a_n (t)$ is the complex amplitude for cavity $n$, and $\omega_0 = (\omega_1 + \omega_2)/2$ is the average mode frequency. The effects of resonance detuning, gain and loss, and inter-cavity couplings are reflected in the dynamics of $a_n(t)$. By substituting Eq. (\ref{eq:CMapprox}) into Eq. (\ref{eq:temporalMaxwell}) and using Eq. (\ref{eq:cavityspatialMaxwell}), we obtain
\begin{align}
	&\sum_{n=1}^{N} \left[ \omega_j^2 \epsilon_{j}({\bf r} - {\bf r}_n) \bm{\Phi}_j({\bf r} - {\bf r}_n) \right] a_n(t) \nonumber \\*
	&= -\epsilon ({\bf r}) \sum_{n=1}^{N} \bm{\Phi}_j({\bf r} - {\bf r}_n) \left[ \left( \partial_t^2 + 2 i \omega_0 \partial_t - \omega_0^2 \right) a_n(t) \right],  \label{eq:slowlyvarying}
\end{align}
where $\partial_t \equiv (\partial/\partial t)$ for simplicity. With the slowly varying approximation, we neglect $\partial_t^2 a_n(t)$ in Eq. (\ref{eq:slowlyvarying}). Subsequently, $\epsilon ({\bf r})$ of the system [Eq. (\ref{eq:permittivity})] is deformed as
\begin{eqnarray}
	\epsilon ({\bf r}) = \epsilon_{j}({\bf r} - {\bf r}_n) + \Psi_{n, r} ({\bf r}) + i \epsilon_i ({\bf r}), \label{eq:permittivity2} \\
	\Psi_{n, r} ({\bf r}) = \sum_{\substack{k=1 \\ k \ne n}}^{N} \left(\epsilon_{k,r} - \epsilon_{e,r} \right) \zeta_{k}({\bf r}), \label{eq:Psi}\\
	i \epsilon_i ({\bf r}) = i \left[ \epsilon_{e,i} + \sum_{k=1}^{N} (\epsilon_{k,i} - \epsilon_{e,i}) \zeta_{k}({\bf r}) \right] \label{eq:permittivityimag}
\end{eqnarray}
so that the terms with $a_n(t)$ can be organized. Here, $\Psi_{n, r} ({\bf r})$ denotes the relative real permittivity profile, excluding the contribution of cavity $n$, and $\epsilon_i ({\bf r}) $ is the entire imaginary potential distribution. Equation (\ref{eq:slowlyvarying}) reduces to
\begin{eqnarray}
	&&\sum_{n=1}^{N} \left[ 2 \delta_j \epsilon_{j}({\bf r} - {\bf r}_n) \bm{\Phi}_j({\bf r} - {\bf r}_n) \right] a_n(t) \nonumber \\*
	&&- \sum_{n=1}^{N} \left[ \omega_0 \Psi_{n, r} ({\bf r}) \bm{\Phi}_j({\bf r} - {\bf r}_n) \right] a_n(t) \nonumber \\
	&&- i \sum_{n=1}^{N} \left[ \omega_0 \epsilon_i ({\bf r}) \bm{\Phi}_j({\bf r} - {\bf r}_n) \right] a_n(t) \nonumber \\
	&&= - 2 i \epsilon ({\bf r}) \sum_{n=1}^{N} \bm{\Phi}_j({\bf r} - {\bf r}_n) \partial_t a_n(t),  \label{eq:fieldexpansion}
\end{eqnarray}
where $\delta_j = \omega_j - \omega_0$, and we have used $\omega_j^2 - \omega_0^2 \sim 2 \omega_0 \delta_j$ based on $|\omega_j - \omega_0| \ll \omega_0$. As shown below, the first, second, and third terms of the left side of Eq. (\ref{eq:fieldexpansion}) contribute to the cavity detuning, real coupling, and on-site gain and loss together with the imaginary coupling, respectively.

Equation (\ref{eq:fieldexpansion}) is now integrated with $\int d{\bf r} \bm{\Phi}_{l} ({\bf r} - {\bf r}_m)$ over the space, with $l$ being the modular index for the $m$th cavity, namely
\begin{equation}
	l =
	\begin{cases}
		1 & (m \ \textrm{mod} \ 2 = 1),\\
		2 & (m \ \textrm{mod} \ 2 = 0).\label{eq:definition_l}
	\end{cases}
\end{equation}
Here, the fields are normalized with $\int d{\bf r} \bm{\Phi}_{l}^{*}({\bf r} - {\bf r}_m) \cdot \epsilon_{l}({\bf r} - {\bf r}_m) \bm{\Phi}_{l}({\bf r} - {\bf r}_m) = \int d{\bf r} \epsilon_{l}({\bf r} - {\bf r}_m) \bm{\Phi}_{l}^2({\bf r} - {\bf r}_m) = 1$, recalling that $\{ \bm{\Phi}_{l=1,2}({\bf r}) \}$ are real. In contrast, we ignore $\delta_j \int d{\bf r} \bm{\Phi}_{l}({\bf r} - {\bf r}_m) \cdot \epsilon_{j}({\bf r} - {\bf r}_n) \bm{\Phi}_j({\bf r} - {\bf r}_n)$ for $m \ne n$, because of $|\delta_j| \ll \omega_0$ and $\int d{\bf r} \bm{\Phi}_{l}({\bf r} - {\bf r}_m) \cdot \epsilon_{j}({\bf r} - {\bf r}_n) \bm{\Phi}_j({\bf r} - {\bf r}_n) \ll 1$. Note that $\epsilon_{l}({\bf r} - {\bf r}_m)$ is based on Eq. (\ref{eq:cavitypermittivity}) for $j \rightarrow l$ and defines medium $l$ at ${\bf r} = {\bf r}_m$ by $\zeta_{0, j}({\bf r} - {\bf r}_m) = \zeta_{m}({\bf r})$.

Because $\partial_t a_n(t)$ is also a small first derivative $\left[\ll \omega_0 a_n(t) \ \textrm{mostly} \right]$, we drop all the tiny modulation on its coefficient coming from other than $\epsilon_{l}({\bf r} - {\bf r}_m)$ in $\epsilon ({\bf r})$, i.e., $\int d{\bf r} \epsilon ({\bf r}) \bm{\Phi}_{l}({\bf r} - {\bf r}_m) \cdot \sum_{n=1}^{N} \bm{\Phi}_j({\bf r} - {\bf r}_n) \partial_t a_n(t) \sim \partial_t a_m(t)$ [see Eqs. (\ref{eq:permittivity2})-(\ref{eq:permittivityimag}) and Ref. \onlinecite{Poon2007} for example]. On the other hand, we keep the inter-cavity overlap integral with the imaginary permittivity, $\omega_0 \int d{\bf r} \epsilon_i ({\bf r}) \bm{\Phi}_{l}({\bf r} - {\bf r}_m) \cdot \sum_{n=1}^{N} \bm{\Phi}_j({\bf r} - {\bf r}_n) a_n(t)$ for $m \ne n$ in the left side of Eq. (\ref{eq:fieldexpansion}), in order to examine how they affect the responses of this non-Hermitian system.

When we limit the couplings to the nearest-neighbor (NN) components (tight-binding approximation), we obtain an equation of motion for $\{ a_m (t)\}$:
\begin{align}
	-i \frac{\textrm{d}a_m}{\textrm{d}t} = \left( \delta_l - i \gamma_l \right) a_m &- \left( \kappa_{m-1, m} + i \kappa'_{m-1, m} \right) a_{m-1} \nonumber \\
	&- \left( \kappa_{m+1, m} + i \kappa'_{m+1, m} \right) a_{m+1}, \quad  \label{eq:CME1}
\end{align}
with the following parameters
\begin{align}
	\delta_l &= \omega_l - \omega_0,\\
	\gamma_l &= \frac{\omega_0}{2} \int d{\bf r} \epsilon_i ({\bf r}) \bm{\Phi}_{l}^2 ({\bf r} - {\bf r}_m) \nonumber \\*
	&\approx \frac{\omega_0}{2} \int d{\bf r} \left[ \epsilon_{e,i} + (\epsilon_{m,i} - \epsilon_{e,i}) \zeta_{m}({\bf r}) \right] \bm{\Phi}_{l}^2 ({\bf r} - {\bf r}_m), \label{eq:gainloss}
\end{align}
\begin{equation}
	\kappa_{m \pm 1, m} = \frac{\omega_0}{2} \int d{\bf r} \Psi_{m \pm 1, r} ({\bf r}) \bm{\Phi}_{l}({\bf r} - {\bf r}_{m}) \cdot \bm{\Phi}_{\bar{l}}({\bf r} - {\bf r}_{m \pm 1}), \label{eq:realcoupling1}
\end{equation}
\begin{equation}
	\kappa'_{m \pm 1, m} = \frac{\omega_0}{2} \int d{\bf r} \epsilon_i ({\bf r}) \bm{\Phi}_{l}({\bf r} - {\bf r}_{m}) \cdot \bm{\Phi}_{\bar{l}}({\bf r} - {\bf r}_{m \pm 1}). \label{eq:imaginarycoupling1}
\end{equation}
Here, $\bar{l}$ is the modular index for the $m \pm 1$th cavity, which is paired with $l$ so that $(l, \bar{l}) = (1,2), \ (2,1)$. Equation (\ref{eq:gainloss}) suggests that the on-site gain or loss $\gamma_l$ in the coupled system can be approximated as that applied to the solitary cavity $l$ with $\bm{\Phi}_{l} ({\bf r})$. $\kappa_{m \pm 1, m}$ and $\kappa'_{m \pm 1, m}$ denote the real and imaginary coupling components, respectively. As shown in Eqs. (\ref{eq:cavityspatialMaxwell}), (\ref{eq:permittivity2}), (\ref{eq:Psi}), and (\ref{eq:fieldexpansion}), we measure the perturbation from the system of the isolated resonances with high and close frequencies $\{\omega_j\}$ based on $\{\epsilon_j ({\bf r} - {\bf r}_{n})\}$. As such, the derived real couplings $\kappa_{m \pm 1, m}$ just evaluate the fields incoming to cavity $m$ and exclude the effect of those inside the adjacent cavities via $\Psi_{m \pm 1, r} ({\bf r})$ [Fig. \ref{fig:system}(c)]. In contrast, the imaginary couplings $\kappa'_{m \pm 1, m}$ involves the entire contribution of the imaginary permittivity distribution $\epsilon_i ({\bf r})$, since the original basis $\{\bm{\Phi}_{j}\}$ is free from it. The latter is hence affected by the gain and loss for both the $m$th and $m \pm 1$th cavity modes. Note that $\kappa_{m,m}/\delta_l \approx 0$ because $\Psi_{m} ({\bf r})$ in $\kappa_{m,m}$ filters out the intensity in cavity $m$ and the cladding material.

Remarkably, we notice that the inversion with reference to ${\bf r}_{m}$, i.e., ${\bf r} - {\bf r}_{m} \rightarrow \bar{\bf r} - {\bf r}_{m} $ and ${\bf r} - {\bf r}_{m \pm 1} \rightarrow \bar{\bf r} - {\bf r}_{m \mp 1} $, links the pairwise coupling factors in both Eqs. (\ref{eq:realcoupling1}) and (\ref{eq:imaginarycoupling1}). This is because $\bm{\Phi}_{\bar{l}}({\bf r} - {\bf r}_{m \pm 1})$ are based on the same cavity mode $\bm{\Phi}_{\bar{l}}({\bf r})$ and because $\bm{\Phi}_{l}({\bf r} - {\bf r}_{m})$ and $\bm{\Phi}_{\bar{l}}({\bf r} - {\bf r}_{m \pm 1})$ have the same parity. We then have
\begin{eqnarray}
	\kappa_{m - 1, m} &=& \frac{\omega_0}{2} \int d \bar{\bf r} \Psi_{m - 1, r} (\bar{\bf r}) \bm{\Phi}_{l}(\bar{\bf r} - {\bf r}_{m}) \cdot \bm{\Phi}_{\bar{l}}(\bar{\bf r} - {\bf r}_{m - 1}) \nonumber \\
	&=&\kappa_{m + 1, m}, \label{eq:realcoupling2}  \\
	\kappa'_{m + 1, m} &=& \kappa'_{m - 1, m}. \label{eq:imaginarycoupling2}
\end{eqnarray}
Here, we have considered the condition that the field products $\bm{\Phi}_{l}({\bf r} - {\bf r}_{m}) \cdot \bm{\Phi}_{\bar{l}}({\bf r} - {\bf r}_{m \pm 1})$ are well confined within the correspondent adjacent cavities, and that the cavities are periodically aligned.

In addition, $\epsilon_i ({\bf r})$ is periodic and hence ensures that the inter-cavity imaginary couplings $(\kappa'_{m - 1, m},\ \kappa'_{m, m + 1})$ in the \textit{forward} direction are equivalent to the \textit{backward} counterparts $(\kappa'_{m, m - 1},\ \kappa'_{m + 1, m})$, namely
\begin{eqnarray}
	\kappa'_{m, m \pm 1} &=& \frac{\omega_0}{2} \int d{\bf r} \epsilon_i ({\bf r}) \bm{\Phi}_{\bar{l}}({\bf r} - {\bf r}_{m \pm 1}) \cdot \bm{\Phi}_{l}({\bf r} - {\bf r}_{m}) \nonumber \\
	&=& \kappa'_{m \pm 1, m}. \label{eq:imaginarycoupling3}
\end{eqnarray}
This is not straightforward for the real couplings. However, the correspondent pairwise factors are approximately equal at least, because the basis modes $(\bm{\Phi}_{l}, \bm{\Phi}_{\bar{l}})$ are nearly identical and the cavities have the common interval $|\vec{r_p}|$:
\begin{eqnarray}
	\kappa_{m, m \pm 1} &=& \frac{\omega_0}{2} \int d{\bf r} \Psi_{m, r}({\bf r}) \bm{\Phi}_{\bar{l}}({\bf r} - {\bf r}_{m \pm 1}) \cdot \bm{\Phi}_{l}({\bf r} - {\bf r}_{m}) \nonumber \\
	&\approx& \kappa_{m \pm 1, m}. \label{eq:realcoupling3}
\end{eqnarray}
Importantly, Eq. (\ref{eq:realcoupling3}) holds equality, $\kappa_{m, m \pm 1} = \kappa_{m \pm 1, m}$, for the cavities of the same structure, i.e., $(w_{x, 1}, w_{y, 1}, w_{z, 1}) = (w_{x, 2}, w_{y, 2}, w_{z, 2})$ and $\epsilon_{1,r} = \epsilon_{2,r}$; see also Eqs. (\ref{eq:Psi}) and (\ref{eq:realcoupling1}). Equation (\ref{eq:imaginarycoupling3}) and (\ref{eq:realcoupling3}) indicate that the coefficient matrix for the mode amplitudes $\{ a_m (t)\}$ in Eq. (\ref{eq:CME1}) is complex-symmetric.

With Eqs. (\ref{eq:CME1})-(\ref{eq:realcoupling3}), we eventually reach the non-Hermitian CMEs for the system,
\begin{align}
	&-i \frac{\textrm{d}a_{2 h - 1}}{\textrm{d}t} = \left( \delta - i \gamma_1 \right) a_{2 h - 1} - \left( \kappa_{r} + i \kappa_{i} \right)\left(a_{2 h - 2} + a_{2 h} \right), \nonumber \\ 
	&-i \frac{\textrm{d}a_{2 h}}{\textrm{d}t} = \left( -\delta - i \gamma_2 \right) a_{2 h} - \left( \kappa_{r} + i \kappa_{i} \right)\left( a_{2 h - 1} + a_{2 h + 1} \right), \quad  \label{eq:CME2}
\end{align}
where $h = 1, 2, \cdots, N/2$, $a_0 = a_{N+1} = 0$. $\kappa_{r} \in \mathbb{R}$ and $\kappa_{i} \in \mathbb{R}$ are the real and imaginary couplings, respectively. $\pm \delta = \pm (\omega_1 - \omega_2)/2$ are the resonance detunings of the cavities from $\omega_0 = (\omega_1 + \omega_2)/2$. $\gamma_l > 0$ and $\gamma_l < 0$ denote gain and loss for cavity $l$. Applying the periodic boundary condition, we have the linear equation $\hat{H}(k) \vec{a_{\rm B}}(k) = \Delta \omega (k) \vec{a_{\rm B}}(k)$ for the system's Bloch eigenvector $\vec{a_{\rm B}}$ and eigen-detuning $\Delta \omega (k)$ with the Hamiltonian
\begin{align}
	&\hat{H}(k) = \nonumber \\ 
	&\left(
	\begin{array}{cc}
		\delta - i \gamma_1 & - (\kappa_{r} + i \kappa_{i}) \left( 1 + e^{-i k L} \right) \\ 
		- (\kappa_{r} + i \kappa_{i}) \left( 1 + e^{i k L} \right) & -\delta - i \gamma_2
	\end{array}\right),
\end{align}
where $k$ is the Bloch wave vector, and $L = 2|\vec{r_p}|$ is the lattice constant for the unit cells. Its solutions can be written as
\begin{align}
	\Delta \omega (k) =& - i \frac{\gamma_1 + \gamma_2}{2} \nonumber \\
	&\pm \sqrt{4 \left(\kappa_{r} + i \kappa_{i}\right)^2 \cos ^2 \frac{k L}{2} - \bigg(\frac{\gamma_1 - \gamma_2}{2} + i \delta \bigg)^2 }, \quad \label{eq:bulkeigenvalue1} \\
	\vec{a_{\rm B}}(k) =& \left( \Delta \omega (k) + \delta + i \gamma_2, \quad -( \kappa_{r} + i \kappa_{i} ) ( 1 + e^{i k L} ) \right)^{\rm T}, \quad \label{eq:bulkeigenvector}
\end{align}
where T denotes transposition.

\subsection{Imaginary coupling and bulk EPs}
The derivation of our model provides useful knowledge about the system response. First, Eqs. (\ref{eq:imaginarycoupling3}) and (\ref{eq:realcoupling3}) mean that the effective system Hamiltonian with an open boundary condition has a complex symmetric form and is hence non-Hermitian. Forward and backward inter-cavity couplings $(c_{m, m + 1}, c_{m + 1, m})$ in non-Hermitian systems just indicate $c_{m, m + 1} \ne c_{m + 1, m}^*$ in general and are affected by how to take the basis. Here, we have resolved such an arbitrary property by determining the nontrivial constraint, $c_{m, m + 1} = c_{m + 1, m} = \kappa_r + i \kappa_i$, with the on-site cavity mode basis. This results in a limited number of independent parameters in the model and hence enables us to figure out the behavior of realistic photonic devices, as seen in Sec. \ref{sec:BHsystem}.

Next, we can estimate how the imaginary couplings depend on the material properties. Let us consider a cavity array composed of identical heterostructures with a real part of their permittivity of $\epsilon_{c,r}$. As the cladding material in such a system is typically a passive dielectric or air, we can safely put $\epsilon_{e,i} = 0$. In this case, $\kappa_i$ is determined by the overlap integral of the fields just within the cavity media, and the cavity mode can be denoted as $\bm{\Phi}_{l}({\bf r}) = \bm{\Phi}_{\bar{l}}({\bf r}) \equiv \bm{\Phi}({\bf r})$. With Eqs. (\ref{eq:permittivityimag}) and (\ref{eq:imaginarycoupling1}), we then have
\begin{eqnarray}
	\kappa'_{m \pm 1, m} &=& \frac{\omega_0}{2} \int d{\bf r} \epsilon_{i}({\bf r}) \bm{\Phi}_{l}({\bf r} - {\bf r}_{m}) \cdot \bm{\Phi}_{\bar{l}}({\bf r} - {\bf r}_{m \pm 1}) \nonumber \\
	&\approx& \frac{\omega_0}{2} \epsilon_{l,i} \int d{\bf r} \zeta_{m}({\bf r}) \bm{\Phi}({\bf r} - {\bf r}_{m}) \cdot \bm{\Phi}({\bf r} - {\bf r}_{m \pm 1}) \nonumber \\
	&& + \frac{\omega_0}{2} \epsilon_{\bar{l},i} \int d{\bf r} \zeta_{m \pm 1}({\bf r}) \bm{\Phi}({\bf r} - {\bf r}_{m}) \cdot \bm{\Phi}({\bf r} - {\bf r}_{m \pm 1}) \nonumber \\
	&=& \frac{\omega_0}{2} (\epsilon_{1,i} + \epsilon_{2,i}) C, \label{eq:imaginarycoupling4}
\end{eqnarray}
where
\begin{eqnarray}
	C &=& \int d{\bf r} \zeta_{m}({\bf r}) \bm{\Phi}({\bf r} - {\bf r}_{m}) \cdot \bm{\Phi}({\bf r} - {\bf r}_{m \pm 1}) \nonumber \\
	&=& \int d{\bf r} \zeta_{m \pm 1}({\bf r}) \bm{\Phi}({\bf r} - {\bf r}_{m}) \cdot \bm{\Phi}({\bf r} - {\bf r}_{m \pm 1}). \label{eq:proportionalconstant}
\end{eqnarray}
Equation (\ref{eq:imaginarycoupling4}) offers an important conclusion that the imaginary coupling is proportional to the sum of the imaginary dielectric constants of the cavities involved. This means that loss-biased systems, even without any real potential contrast (i.e., $\epsilon_{1,i} + \epsilon_{2,i}> 0 $, $\delta = 0$), will miss the exact EP, because $\kappa_{i}$ results in a finite imaginary component inside the radical term in $\Delta \omega (k)$; see Eq. (\ref{eq:bulkeigenvalue1}). Moreover, the local overlap integral $C$ also appears in the expression for the real part of the coupling, i.e., Eq. (\ref{eq:realcoupling1}). As a result, the order of the ratio between the real and imaginary couplings is given by
\begin{equation}
	\frac{\kappa_i}{\kappa_r} \sim \frac{\epsilon_{1,i} + \epsilon_{2,i}}{\epsilon_{c,r} - \epsilon_{e,r}}. \label{eq:ratiocoupling}
\end{equation}
Equation (\ref{eq:ratiocoupling}) shows why $\kappa_i$ is second-order in terms of the NN mode overlap integral. In the context of the EP formation requiring the balance between on-site imaginary potential contrast and real couplings, the right-hand side of Eq. (\ref{eq:ratiocoupling}) implicitly reflects the overlap integral compared to the on-site field intensity. The approximate equality here means that most on-chip cavities are planar devices and hence governed by effective indices, causing a deviation from Eq. (\ref{eq:ratiocoupling}) based on the material permittivities.

With Eq. (\ref{eq:bulkeigenvalue1}), we obtain a sufficient condition for $\Delta \omega (k)$ to have EPs as
\begin{equation}
	\frac{1}{2} \frac{g + i \delta}{\kappa_r + i \kappa_i} = \eta \in \mathbb{R}, \quad |\eta| \le 1. \label{eq:EPCondition1}
\end{equation}
where $g = (\gamma_1 - \gamma_2)/2$. Equation (\ref{eq:EPCondition1}) requires that $g + i\delta$ and $\kappa_r + i \kappa_i$ are parallel in the complex plane. Such a $\delta$ appears to be achievable for each $g$ when $|g| \le 2|\kappa_r|$. However, both $g$ and $\kappa_i$ have been clarified to depend on the imaginary permittivities $(\epsilon_{1,i}, \epsilon_{2,i})$ with Eqs. (\ref{eq:gainloss}) and (\ref{eq:imaginarycoupling4}), and the desired argument ${\rm Arg}(g + i\delta)$ is hence not constant for active devices. If the detuning $\delta$ also varies significantly with external pumping, via thermal and carrier effects, it might be generally difficult to reach an EP in experiment.

To address this problem, we show in Sec. \ref{sec:BHsystem} that the permittivity-induced imaginary coupling is suppressed when the system has balanced gain and loss, i.e., $\kappa_i = 0$ for $\epsilon_{1,i} = - \epsilon_{2,i}$, in a simulation of buried-heterostructure photonic crystal nanocavities.

\section{\label{sec:CMTRad} Non-Hermitian coupled-mode theory: impact of radiation}
\subsection{Modeling of radiation in coupled-mode theory}
Every single mode of practical cavities and waveguides exhibits radiation loss, which has a clearly different physical origin from the material absorption studied in the last section. Here, we show a way to treat the radiation within the CMT framework and discuss its impact on the imaginary coupling.

We first consider a dielectric cavity medium, a cladding material of finite size, and an extensive air layer outside for each of the unperturbed single-mode cavities. All the materials here have real dielectric constants, and the cavity modes are supposed to have strong light confinement. A minute part of each mode leaks out of the cladding layer and couples with radiative plane waves. Here, we consider enclosing the whole system with a perfect electric conductor (PEC) located at infinity, so that the weak radiation fields are kept within the defined air domain. In this case, the operator $\nabla \times \epsilon^{-1} ({\bf r}) \nabla \times$ of the wave equation for the magnetic fields is Hermitian. Moreover, an eigenstate for the Maxwell equations has magnetic and electric fields with a common eigenfrequency. Thus, we can take a series of real electric cavity modes $\{\bm{\Phi}_{j}({\bf r}) \in \mathbb{R}^3 \}$ and their frequencies $\{\omega_j \in \mathbb{R} \}$ for our basis:
\begin{align}
	\nabla \times \nabla \times \bm{\Phi}_j({\bf r}) =& \frac{\omega_j^2}{c^2} \epsilon'_{j}({\bf r})  \bm{\Phi}_j({\bf r}), \label{eq:cavityspatialMaxwell2} \\
	\epsilon'_j ({\bf r}) = 1 \cdot \zeta_{\rm A}({\bf r}) + \epsilon_{\rm e,r} \zeta_{\rm B}({\bf r}) &+ (\epsilon_{ j,{\rm r}} - \epsilon_{\rm e,r}) \zeta_{0,j}({\bf r}), \label{eq:cavitypermittivity2}
\end{align}
where $\epsilon'_j ({\bf r})$ is the permittivity distribution for the $j$th isolated cavity system, and $\bm{n} \times \bm{\Phi}_j({\bf r}) = \bm{0} \ (r \rightarrow \infty)$ with $\bm{n}$ being a unit vector normal to the PEC. $\zeta_{0,j}$, $\zeta_{\rm A}({\bf r})$ and $\zeta_{\rm B}({\bf r})$ are products of Heaviside functions that mark the position of the cavity medium, air layer, and cladding material, respectively; see Eq. (\ref{eq:cavityzeta}) again for example. The permittivities of the cavity, cladding material, and air are $\epsilon_{j,{\rm r}}$, $\epsilon_{\rm e,r}$ and $\epsilon_{\rm A} = 1$, with $1 < \epsilon_{\rm e,r} < \epsilon_{ j,{\rm r}}$ satisfied for the formation of cavity modes. The reason for assuming such a basis is that optical modes with radiation fields cannot be expanded rigorously by confined modes just with exponentially decaying tails, which are used in Sec. \ref{sec:CMT}.

\begin{figure}[t!]
	\includegraphics[width=6.0cm]{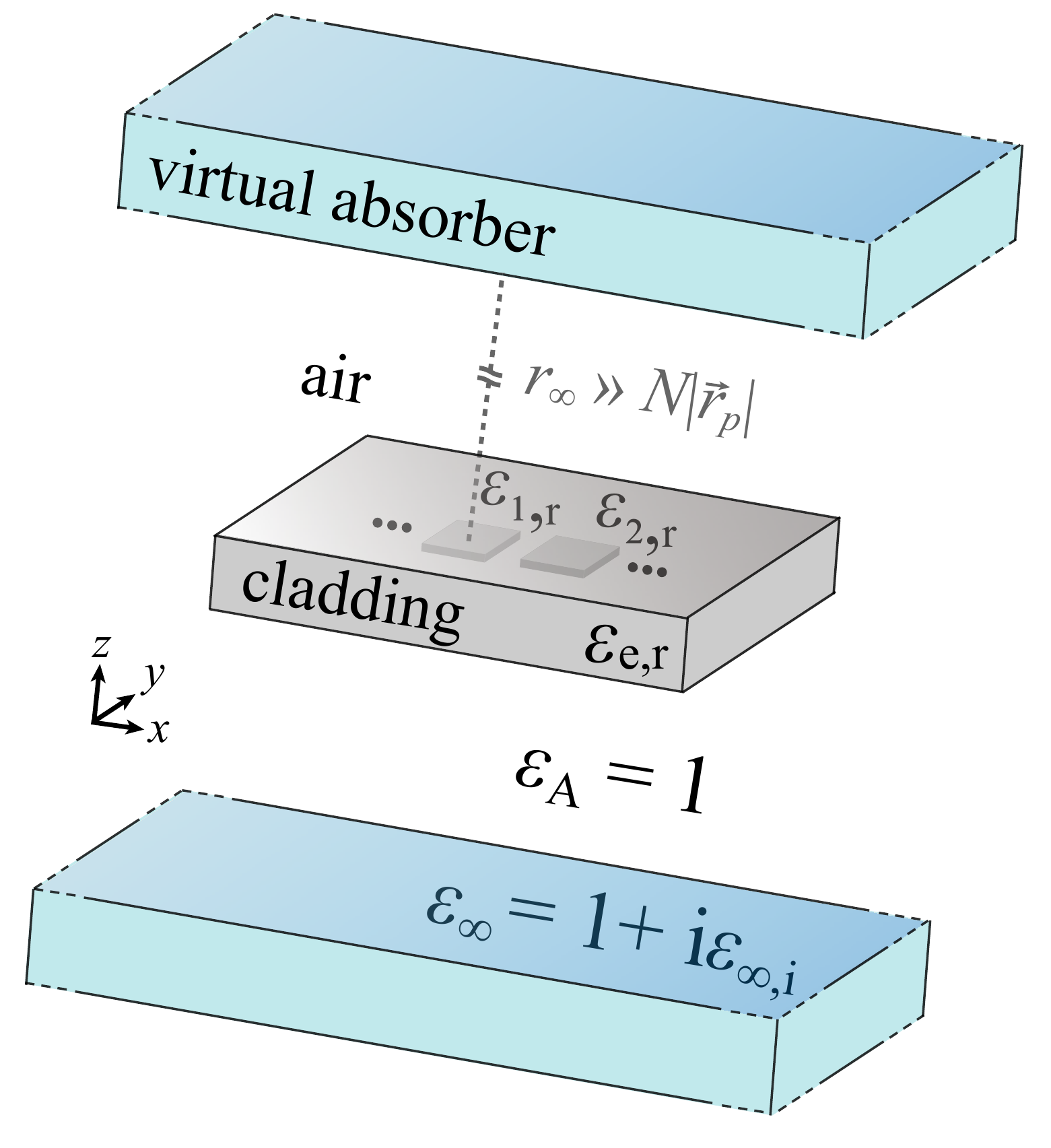}
	\caption{\label{fig:Radiation} Modeling radiation loss as a virtual absorbing layer placed at infinity and surrounds a cavity array. The system originally has $N$ dielectric cavity media, a cladding material layer with finite size, and a boundless air region outside. We introduce an imaginary permittivity $i\epsilon_{\infty, i}$ to a part of the air domain and reckon this as a virtual absorber that totally damps the light. We assume that the minimum distance $r_{\infty}$ between each cavity and the absorbing layer is orders of magnitude larger than the array length $N|\vec{r_p}|$. Although the absorber is supposed to be distributed far in $\pm x$ and $\pm y$ directions, it is omitted for a better view of the system structure.}
\end{figure}

Here, we model the radiation effects by placing a virtual absorbing layer just in front of the PEC at infinity (Fig. \ref{fig:Radiation}), in analogy to well-known simulation techniques such as the finite-difference time-domain (FDTD) method and finite element method (FEM). Although the reflection from such an absorber can be totally suppressed \cite{Berenger1994}, we need a magnetic conductivity that introduces an additional term to the temporal Maxwell equation for that case. For simplicity, we just suppose that its permittivity $\epsilon_{\infty}$ has the same real part as that of air, ${\rm Re} \ \epsilon_{\infty} = 1$, and a small imaginary part ${\rm Im} \ \epsilon_{\infty} = \epsilon_{\infty, i}$, and that its size is large enough to damp the light thoroughly. Because we can make this virtual layer thicker at will, we should be able to find a value of $\epsilon_{\infty, i}$ with which its reflection does not affect the response of the cavities. The location of the absorber is denoted as $\zeta_{\infty}({\bf r})$ and expressed with some combination of Heaviside functions, so that its relative permittivity distribution is written as $\epsilon_{\infty}({\bf r}) \equiv [1 + i\epsilon_{\infty, i} ]\zeta_{\infty}({\bf r})$. We also put $\zeta_{\infty}({\bf r}) \subset \zeta_{\rm A}({\bf r})$, and can hence regard the perturbation by $\epsilon_{\infty}({\bf r})$ as adding $i\epsilon_{\infty, i}$ to a part of the original air layer $\zeta_{\rm A}({\bf r})$ in Eq. (\ref{eq:cavitypermittivity2}). 
An open system with $N$ cavities ceases to be affected by the PEC in this arrangement, and its permittivity distribution $\epsilon' ({\bf r})$ has apparent non-Hermiticity just by the absorption $i\epsilon_{\infty, i} \zeta_{\infty}({\bf r})$ in the virtual layer, namely
\begin{align}
	&\epsilon' ({\bf r}) = \epsilon_{\rm R} ({\bf r}) + i \epsilon_{\infty, i} \zeta_{\infty}({\bf r}), \label{eq:PALpermittivity} \\
	&\epsilon_{\rm R} ({\bf r}) = 1 \cdot \zeta_{\rm A}({\bf r}) + \epsilon_{\rm e,r} \zeta_{\rm B}({\bf r}) + \sum_{n = 1}^{N} (\epsilon_{ n,{\rm r}} - \epsilon_{\rm e,r}) \zeta_{n}({\bf r}), \label{eq:PALpermittivityreal}
\end{align}
where the $n$th cavity is again defined with $\zeta_{n}({\bf r})$ shown in Eq. (\ref{eq:zeta}) and located inside the background cladding region, $\zeta_{n}({\bf r}) \subset \zeta_{\rm B}({\bf r})$.

We again consider an array of unit cells with two distinct cavities $(l = 1, 2)$, where they have contrast of radiation loss arising from small structural modulation, $(w_{x, 1}, w_{y, 1}, w_{z, 1}) \ne (w_{x, 2}, w_{y, 2}, w_{z, 2})$. By following the same derivation as in Sec. \ref{sec:CMT}, we obtain the CMEs for the system with $\epsilon' ({\bf r})$ as
\begin{align}
	&-i \frac{\textrm{d}a_{2 h - 1}}{\textrm{d}t} = \left( \delta - i \Gamma_1 \right) a_{2 h - 1} - \left( \kappa_{r} + i K_{i} \right)\left(a_{2 h - 2} + a_{2 h} \right), \nonumber \\ 
	&-i \frac{\textrm{d}a_{2 h}}{\textrm{d}t} = \left( -\delta - i \Gamma_2 \right) a_{2 h} - \left( \kappa_{r} + i K_{i} \right)\left( a_{2 h - 1} + a_{2 h + 1} \right), \label{eq:CME3}
\end{align}
while the on-site radiation loss $\Gamma_{m} < 0$ and imaginary coupling $K_{i}$ within the NN read
\begin{align}
	\Gamma_l &\approx \frac{\omega_0}{2} \int d{\bf r} \epsilon_{\infty, i} \zeta_{\infty}({\bf r}) \bm{\Phi}_{l}^2 ({\bf r} - {\bf r}_m), \qquad \label{eq:radiationloss} \\
	K_i &= \frac{\omega_0}{2} \int d{\bf r} \epsilon_{\infty, i} \zeta_{\infty}({\bf r}) \bm{\Phi}_{l}({\bf r} - {\bf r}_{m}) \cdot \bm{\Phi}_{\bar{l}}({\bf r} - {\bf r}_{m \pm 1}) \label{eq:imcouplingradiation},
\end{align}
where $(l, \bar{l}) = (1, 2), (2, 1)$ correspond to the modular indices for $(m, m \pm 1)$. As shown in Eq. (\ref{eq:radiationloss}), the radiation loss of the field is formulated with half of the mode intensity that reaches the distant absorber. The definition of the real coupling $\kappa_{r}$ is shown in Eq. (\ref{eq:realcoupling3}). We have considered that Eq. (\ref{eq:imcouplingradiation}) for all $m$ is based on the single pair of cavity modes $\{\bm{\Phi}_{1}, \bm{\Phi}_{2}\}$ and the virtual absorptive layer is supposed to be equidistant from any of two adjacent cavities. Such an imaginary coupling term might also be derived when we virtually place dissipation ports \cite{Suh2004} so that each of them is symmetrically coupled with each pair of NN cavities. However, our CMT will cover more general cases, where the entire structure may be asymmetric or anisotropic, by describing the coupling terms as integral forms.

Here, we find a remarkable relation:
\begin{align}
	&\frac{\omega_0}{2} \int d{\bf r} \epsilon_{\infty, i} \zeta_{\infty}({\bf r}) \bigg| \frac{1}{\sqrt{2}} \big[ \bm{\Phi}_{l}({\bf r} - {\bf r}_{m}) + \bm{\Phi}_{\bar{l}}({\bf r} - {\bf r}_{m \pm 1}) \big] \bigg|^2 \nonumber \\ 
	&= \frac{\Gamma_1 + \Gamma_2}{2} + K_i \label{eq:sumfieldloss}.
\end{align}
Equation (\ref{eq:sumfieldloss}) denotes radiation loss for \textit{superposition} of the two basis cavity modes, $(\bm{\Phi}_{l}({\bf r} - {\bf r}_{m}) + \bm{\Phi}_{\bar{l}}({\bf r} - {\bf r}_{m \pm 1}))/\sqrt{2}$. Each of the square terms [$\Gamma_1$ and $\Gamma_2$: Eq. (\ref{eq:radiationloss})] is the solitary contribution of each cavity. In contrast, the cross terms $K_i$ correspond to the forward and backward imaginary couplings [Eq. (\ref{eq:imcouplingradiation})] and measure explicitly the impact of the interference between $\bm{\Phi}_{1}$ and $\bm{\Phi}_{2}$. Within the NN approximation, we can extend our discussion to the entire $N$-cavity system, namely
\begin{equation}
	\frac{\omega_0}{2} \int d{\bf r} \epsilon_{\infty, i} \zeta_{\infty}({\bf r}) \bigg| \frac{1}{\sqrt{N}} \sum _{n=1}^N \bm{\Phi}_{l} ({\bf r} - {\bf r}_{n}) \bigg| ^2 
	\approx \frac{\Gamma_1 + \Gamma_2}{2} + 2 K_i \label{eq:Ncavitiesloss}.
\end{equation}
The imaginary coupling $K_i$ is hence an essential factor determining the net radiation loss of coupled resonators. It is notable that the coefficient of $K_i$ in Eq. (\ref{eq:Ncavitiesloss}) is twice as large as that in Eq. (\ref{eq:sumfieldloss}). This difference corresponds to whether each cavity couples with another or two other ones.

Equation (\ref{eq:imcouplingradiation}) does not mention how $\bm{\Phi}_{l}$ and $\bm{\Phi}_{\bar{l}}$ interfere, because it depends on their detailed spatial shapes. $K_i$ will include the contribution of both the interference of evanescent fields residing mostly in the cladding layer and that of plane waves radiated into the air. For the latter, each cavity would behave like a point source and hence implicitly provides an approximate factor of $\propto 1/|{\bf r} - {\bf r}_{n}|$, with ${\bf r}_{n}$ being the cavity position. Such a non-local property may lead to non-negligible second-nearest-neighbor (SNN) radiation-based imaginary couplings, third-nearest-neighbor ones, and so on. They are in principle contained in the left-hand side of Eq. (\ref{eq:Ncavitiesloss}). 

Here, we compare the radiation-induced imaginary couplings and permittivity-based ones. As seen in Eqs. (\ref{eq:CME2}) and (\ref{eq:CME3}), the effect of radiation and that of the imaginary permittivities of gain media appear as the same form in the CMEs. The difference between them is that different cavities can make separable contributions in the latter. Then, what if the cladding material has a finite imaginary permittivity $\epsilon_{\rm e,i}$ instead of the cavities? In fact, that case also yields an imaginary coupling $\kappa_{i, {\rm B}}$ in the symmetric form. $\kappa_{i, {\rm B}}$ is given by the overlap integral of two adjacent cavity modes within the cladding material, which is multiplied by $\epsilon_{\rm e,i}$, and this term is analogous to the radiation-induced imaginary coupling, i.e., Eq. (\ref{eq:imcouplingradiation}). The following interpretation of the imaginary coupling helps us understand such a similarity from the local perspective.

Let us consider that the spatial cavity modes $\{\bm{\Phi}_{j}({\bf r} - {\bf r}_{n}) \in \mathbb{R}^3 \}$ are perturbed by the additional loss in the cladding material and interpret this effect as the modulation of the coupling terms. If the evanescent fields between cavities undergo small leakages or absorption, they will have an extra factor $\exp(- \alpha |\vec{r_p}|)$, where $\alpha$ is the net extinction coefficient. Because we focus on the case for $\alpha |\vec{r_p}| \ll 1$, this term can be decomposed as $\exp(- \alpha |\vec{r_p}|) \approx 1 - \sin (\alpha |\vec{r_p}|) \ [\approx \cos (\alpha |\vec{r_p}|) - \sin (\alpha |\vec{r_p}|)]$. It means that the loss during the coupling almost preserves the original evanescent fields while gives a small sinusoidal (quadrature) factor that induces phase retardation of tunneling waves. Thus, the basis modes overlapping within the cavities come to exhibit the interference that is absent in the Hermitian system, and the resultant change in their intensity (i.e. energy) corresponds to the relative loss in the cladding layer. In the CMT framework, this process is reflected effectively in a finite argument of the complex coupling, and hence $K_i$, regardless of how the fields are damped in the coupling paths.

Nonetheless, we emphasize that the permittivity-induced imaginary couplings are mostly limited to the NN components, because they are based on evanescent fields. In contrast, distant cavities can have finite radiation-based imaginary couplings. Such non-local nature of radiation is expected to be essential in systems with imaginary band structures that significantly deviate from cosinusoidal shapes, one of which is shown later in Fig. \ref{fig:BHbandstructure1}(b).

\subsection{Non-Hermitian two-cavity system}
In Sec. \ref{sec:RBEP}, we examine the impact of the radiation-induced imaginary coupling by simulating a system of two cavities where the CMT is exact in terms of the coupling profile. Here, we describe the theoretical responses of such a system. Note that the analysis is also applicable quantitatively for the amplification- and absorption-based system. Equation (\ref{eq:CME3}) can reduce to
\begin{eqnarray}
	-i \frac{\rm d}{{\rm d}t}
	\left(
	\begin{array}{c}
		a_1 \\ 
		a_2
	\end{array}
	\right) &=&
	\left(
	\begin{array}{cc}
		\delta - i \Gamma_1 & - \left( \kappa_{r} + i K_{i} \right)  \\ 
		- \left( \kappa_{r} + i K_{i} \right) & -\delta - i \Gamma_2
	\end{array}
	\right)
	\left(
	\begin{array}{c}
		a_1 \\ 
		a_2
	\end{array}
	\right) \nonumber \\
	&\equiv& \hat{H}
	\left(
	\begin{array}{c}
		a_1 \\ 
		a_2
	\end{array}
	\right). \label{eq:radiationbasedCMT}
\end{eqnarray}
The eigen-detuning $\Delta \omega_{\pm}$ for Eq. (\ref{eq:radiationbasedCMT}) is given by
\begin{equation}
	\Delta \omega_{\pm} = - i\frac{\Gamma_1 + \Gamma_2}{2}
	\pm \sqrt{ \left(\kappa_{r} + i K_{i}\right)^2 - \bigg(\frac{\Gamma_1 - \Gamma_2}{2} + i \delta \bigg)^2 }. \label{eq:twocaveigenvalue1}
\end{equation}
By a trivial modification of Eq. (\ref{eq:twocaveigenvalue1}) with $l \equiv (\Gamma_1 - \Gamma_2)/2$, we can identify the condition for the EP as $(\kappa_{r} + i K_{i})^2 - (l + i \delta)^2 = 0$, namely
\begin{equation}
	\begin{cases}
		\kappa_{r} = l, \ K_{i} = \delta, \\
		\kappa_{r} = -l, \ K_{i} = -\delta.
	\end{cases} \label{eq:EPCondition}
\end{equation}
The two cases here correspond to two EPs in the entire parameter space. This simply means the equivalence of the two cavities; either of them can be lossier. In addition to the balance between the real coupling $\kappa_{r}$ and loss contrast $l$, a finite imaginary coupling $K_{i}$ must be compensated by the frequency detuning $\delta$ for reaching the EP. This means that EPs in real systems, more or less, should be found in the hybrid of PT-symmetric and anti-PT-symmetric potential in terms of the on-site mode basis, as has been shown phenomenologically \cite{Benisty2012}. 

We further discuss the topological robustness of the EPs in the two-cavity system with the imaginary coupling. Here, we exclude the average potential contribution from the effective Hamiltonian $\hat{H}$, which does not affect the essential behavior of the system. The resultant "unbiased" Hamiltonian is defined as
\begin{equation}
	\hat{H}' \equiv \hat{H} + i \frac{\Gamma_1 - \Gamma_2}{2} \hat{I}  = 
	\left(
	\begin{array}{cc}
		\delta - i l & - \left( \kappa_{r} + i K_{i} \right)  \\ 
		- \left( \kappa_{r} + i K_{i} \right) & -\delta + i l
	\end{array}
	\right),
\end{equation}
where $\hat{I}$ is the $2 \times 2$ identity matrix. By separating the real and imaginary parts of the radical term in Eq. (\ref{eq:twocaveigenvalue1}), we obtain the eigenvalues of $\hat{H}'$ as
\begin{equation}
	\Delta \omega_{\pm}' = \pm \sqrt{\kappa^2_{r} - K^2_{i} - l^2 + \delta^2 + 2i (\kappa_{r} K_{i} - l \delta)}. \label{eq:twocaveigenvalue2}
\end{equation}
Remarkably, we notice a condition
\begin{equation}
	\kappa_{r} K_{i} - l \delta = 0, \label{eq:generalPTcondition}
\end{equation}
for purely real or imaginary eigenvalues $\Delta \omega_{\pm}'$, which are also seen in ideal PT-symmetric coupled cavities. We show below that systems with Eq. (\ref{eq:generalPTcondition}) actually respect a \textit{generalized} PT symmetry \cite{Okugawa2019,Yoshida2019,Kawabata2019}.

By using Pauli matrices $(\hat{\sigma_x}, \hat{\sigma_y}, \hat{\sigma_z})^{\rm T} \equiv \bm{\hat{\sigma}}$, the unbiased Hamiltonian is written as
\begin{eqnarray}
	\hat{H}' &=& -(\kappa_r + i K_i) \hat{\sigma_x} + (\delta - i l) \hat{\sigma_z} \nonumber \\
	&=& (\bm{b} + i \bm{d}) \cdot \bm{\hat{\sigma}} \label{eq:UBspinH},
\end{eqnarray}
where $\bm{b} = (-\kappa_r, 0, \delta)^{\rm T}$ and $\bm{d} = (-K_i, 0, -l)^{\rm T}$. Here, $\hat{H}'$ does not include any $\hat{\sigma_y}$ components, and Eq. (\ref{eq:generalPTcondition}) means the orthogonality of the real and imaginary spin coefficient vectors, $\bm{b} \cdot \bm{d} = 0$. In this case, the orthogonal transformation denoting a rotation of the effective spinor $(a_1, a_2)^{\rm T}$ around $y$ axis,
\begin{eqnarray}
	\hat{R} &\equiv& \exp \left( -i \frac{\theta}{2} \hat{\sigma_y} \right) \nonumber \\
	&=& \left(
	\begin{array}{cc}
		\cos (\theta/2) & \sin (\theta/2)  \\ 
		-\sin (\theta/2) & \cos (\theta/2)
	\end{array}
	\right), \label{eq:rotator} \\
	\theta &=& \tan^{-1}\left( \frac{\delta}{\kappa_r} \right),
\end{eqnarray}
is found to reframe the system so that the modified Hamiltonian has purely real couplings and imaginary on-site potential contrast,
\begin{eqnarray}
	\tilde{H}' = \hat{R} \hat{H}' \hat{R}^{-1} 
	= \left( 
	\begin{array}{cc}
		-i \xi & -\chi  \\ 
		-\chi & i \xi
	\end{array}
	\right), \\
	\xi = \frac{\kappa_r l + K_i \delta}{\sqrt{\kappa_r ^2 + \delta ^2}}, \quad \chi = \sqrt{\kappa_r ^2 + \delta ^2}.
\end{eqnarray}
Here, a tilde is used to mark an operator in the rotated system. It immediately follows that $\tilde{H}'$ respects the conventional PT symmetry
\begin{equation}
	\tilde{P} \tilde{T} \tilde{H}' (\tilde{P} \tilde{T})^{-1} = \tilde{H}', \label{eq:PTsym}
\end{equation}
where $\tilde{P} = \tilde{\sigma_x}$ is the inversion operation, and $\tilde{T} = \tilde{K}$ denotes complex conjugation. Equation (\ref{eq:PTsym}) reduces to the general PT symmetry for the original basis
\begin{equation}
	\hat{U} \hat{H}'^{*} \hat{U}^{-1} = \hat{H}', \quad \hat{U}\hat{U}^{*} = +1, \label{eq:generalPTsym}
\end{equation}
where $\hat{U} = \hat{U}^{*} = \hat{R}^{-1} \tilde{P} \hat{R}$ is a unitary operator.

Our EPs with Eq. (\ref{eq:EPCondition}) satisfy Eq. (\ref{eq:generalPTcondition}). Thus, they are characterized by the general PT symmetry \cite{Kawabata2019}, i.e., Eq. (\ref{eq:generalPTsym}). The complex spectrum $\Delta \omega_{\pm}'$ has fractional rounds of phase vortices around the EPs, which are denoted by $\nu = \pm 1/2$ depending on whether their direction is clockwise or counter-clockwise \cite{Leykam2017,Zhou2018,Takata2018}. This fractional charge $\nu$ comes from the fact that a square-root complex function needs two laps of variables around the branch point to get back to the same value. It actually corresponds to the $\mathbb{Z}_2$ topological invariant of the general PT-symmetric EPs. The EPs with finite $\nu$ are hence robust to continuous changes in the parameters. In other words, for any systems with specific $(\kappa_r, K_i)$, there exists a trajectory [Eq. (\ref{eq:generalPTcondition})] that guarantees the general PT symmetry in the $(l, \delta)$ space. We can always find the EPs incorporated there by varying the two parameters $l$ and $\delta$, unless the EPs "annihilate" each other just at the origin: $\kappa_r = l = 0$, $\kappa_i = \delta = 0$.

\section{\label{sec:BHsystem}Imaginary couplings and exceptional points in coupled active heterostructure nanocavities}
\begin{figure}[t!]
	\includegraphics[width=8.6cm]{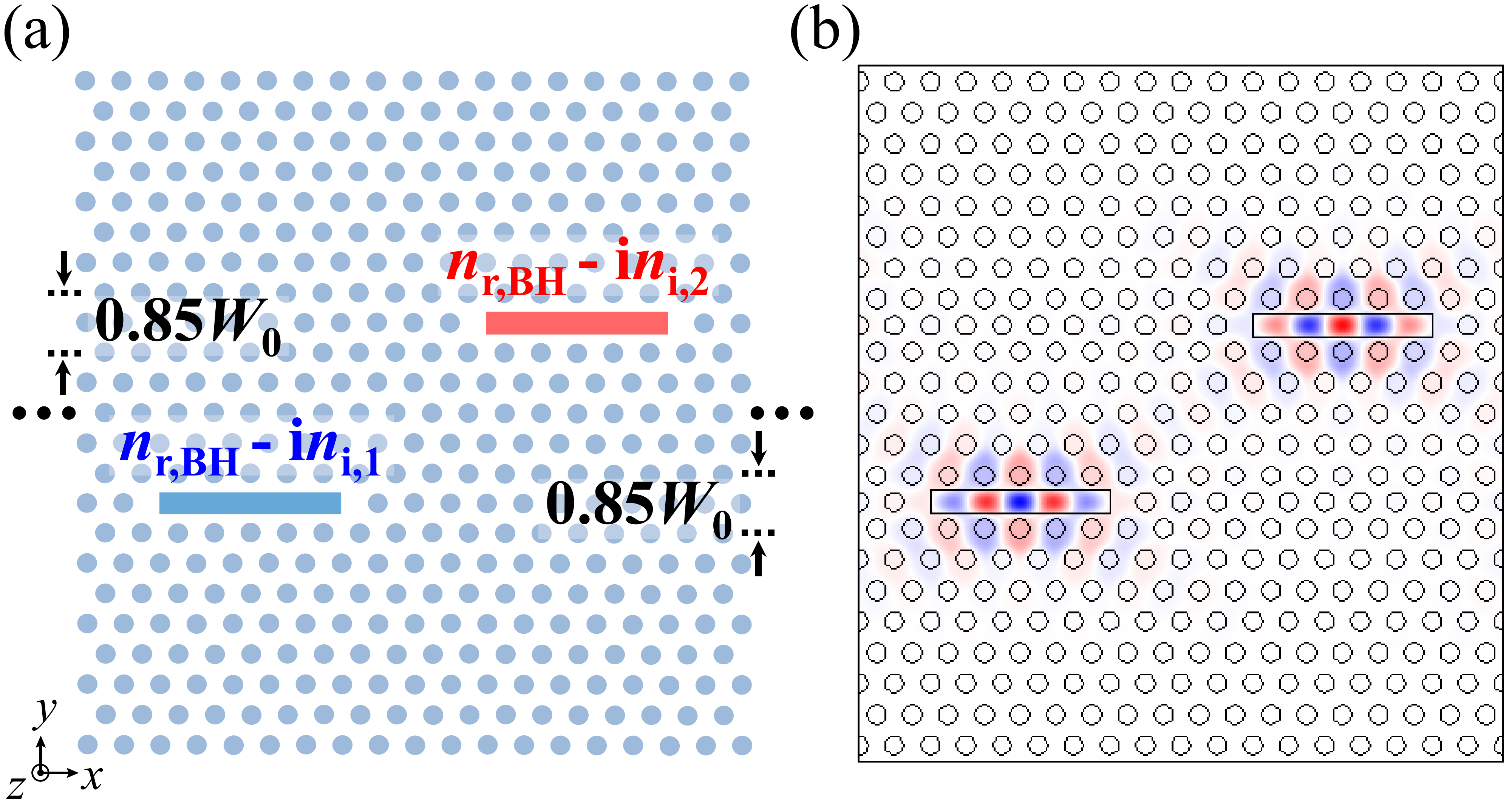}
	\caption{\label{fig:schematicBH} (a) Top-view schematic of the unit cell of the considered nanocavity array. The system is composed of a two-dimensional triangular-lattice photonic crystal slab and periods of two active heterostructures that are buried in narrow line defects and arranged in a zigzag geometry. (b) $z$-component of the magnetic fields of a coupled ground mode for $n_{\rm 1,i} = n_{\rm 2,i} = 0$. Evanescent fields spread in diagonal directions, and the strong coupling is hence achieved.} 
\end{figure}

In this section, we apply the theoretical framework developed in Sec. \ref{sec:CMT} to a practical simulation example of an array of buried-heterostructure photonic crystal nanocavities \cite{Takata2021,Matsuo2010,Takeda2013,Takata2017}, which is one of the well-behaved non-Hermitian coupled-mode platforms. Figure \ref{fig:schematicBH}(a) depicts the top view of its unit cell schematically. The system is constructed on an InP slab ($n_{\rm InP} = 3.16$) with a thickness of $t = 250 \ {\rm nm}$ suspended in the air and has a periodic condition in the horizontal ($x$) direction. The underlying two-dimensional photonic crystal is a triangular lattice of circular air holes with radius $R = 120 \ {\rm nm}$ and lattice constant $a = 450 \ {\rm nm}$. Bulk InGaAsP heterostructures ($n_{\rm BH,r} = 3.539$) with dimensions of $5a (2.25 \ \mu{\rm m}) \times 0.3 \ \mu{\rm m} \times 0.15 \ \mu{\rm m}$ are buried in line defects with five air holes removed. They are arranged in a zigzag alignment and form effectively a one-dimensional coupled cavity chain, because their ground mode has evanescent tails in $\Gamma$-M and $\Gamma$-M' directions, as shown in Fig. \ref{fig:schematicBH}(b). The cavities are all equally spaced, and the unit cell period is $L = 18 a$. The line defects are narrowed by shifting both the upper and lower rows of air holes toward the center so that their width is $0.85 W_0 \ (W_0 = \sqrt{3}a)$. This structural modulation improves the optical confinement of the coupled modes. The single nanocavity with eight layers of photonic crystals on both sides, which are also adopted in the periodic system, has a $Q$ factor of $Q = 1.9 \times 10^5$ for the ground mode. Photonic band structures of the three-dimensional system are computed by a numerical solver based on the finite-element method for different configurations of the imaginary indices of the heterostructures denoted as $n_{\rm 1,i}$ and $n_{\rm 2,i}$.

\begin{figure*}[t!]
	\includegraphics[width=14.0cm]{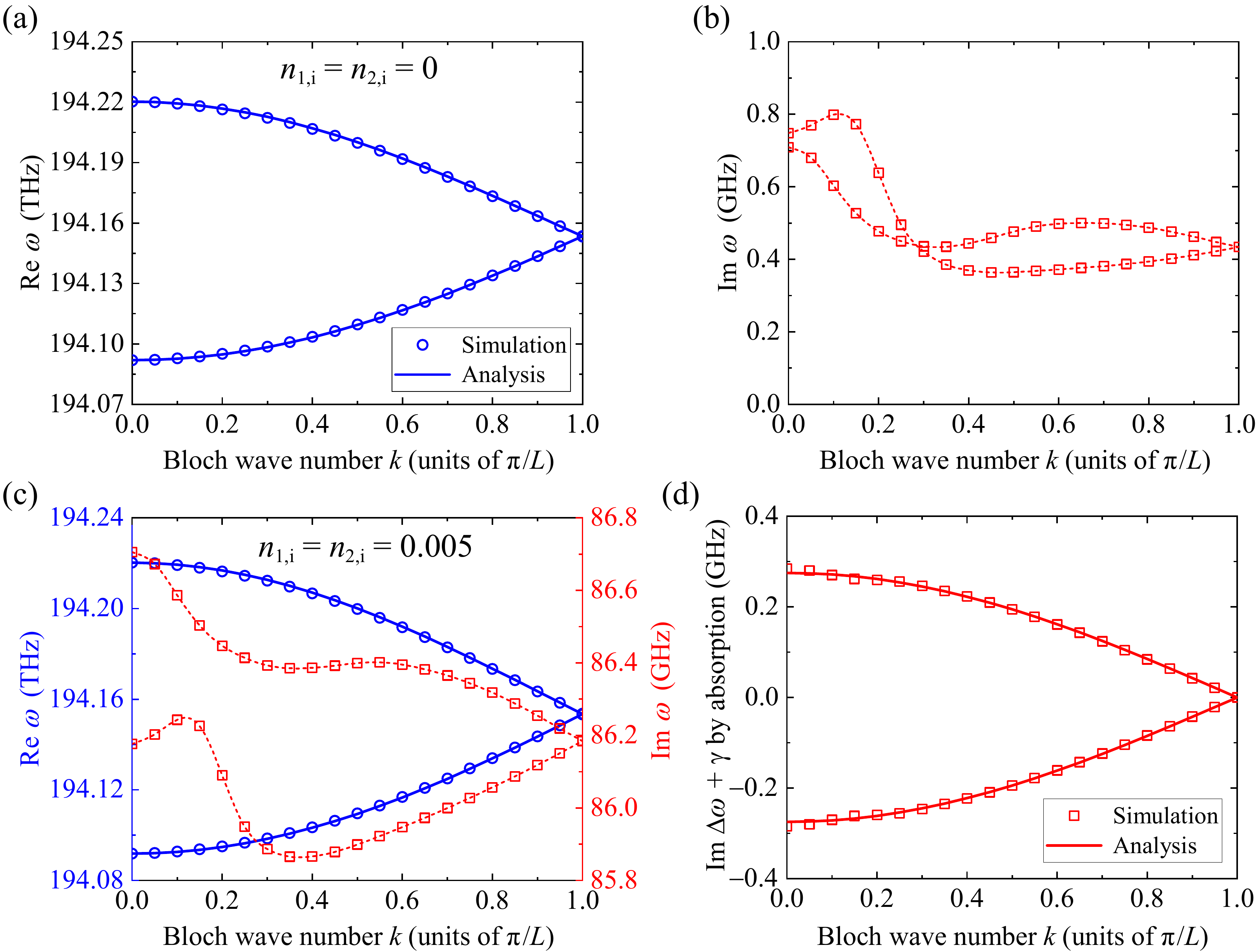}
	\caption{\label{fig:BHbandstructure1} Complex band structures $\omega (k) = \omega_0 + \Delta \omega (k)$ for the system with uniform refractive indices of nanocavities. (a) ${\rm Re} \ \omega (k)$ and (b) ${\rm Im} \ \omega (k)$ for $n_{\rm 1,i} = n_{\rm 2,i} = 0$. Here, ${\rm Im} \ \omega (k)$ in (b) is induced purely by radiation. (c) $\omega (k)$ of an absorptive system with $n_{\rm 1,i} = n_{\rm 2,i} = 0.005$. Data colored blue: ${\rm Re} \ \omega (k)$. Those in red: ${\rm Im} \ \omega (k)$. (d) Contribution of the cavities' material absorption to the imaginary dispersion in (c), which is obtained as the difference between ${\rm Im} \ \omega (k)$ in (c) and that in (b). Its folded cosinusoidal shape confirms the existence of the NN imaginary coupling $\kappa_i$. Markers: result of the finite-element simulation. Solid curves: analytic result. $\kappa_r = 32.08 \ {\rm GHz}$, $\phi = 0.04 \ {\rm rad}$. (a) $\omega_0 = 194.1547 \ {\rm THz}$, $\rho = 0.662 \ {\rm GHz}$ and (c) $\omega_0 = 194.1546 \ {\rm THz}$, $\rho = 0.661 \ {\rm GHz}$ for ${\rm Re} \ \Delta \omega (k)$. $\kappa_i = -0.137 \ {\rm GHz}$ for (d). Dotted curves: guide for the eye.}
\end{figure*}

Figure \ref{fig:BHbandstructure1}(a) and (b) present the complex band structure $\omega (k) \equiv \omega_0 + \Delta \omega (k)$ for the system of cold cavities, i.e., $n_{\rm 1,i} = n_{\rm 2,i} = 0$.
As seen in the analytic eigen-detuning [$\Delta \omega (k)$ in Eq. (\ref{eq:bulkeigenvalue1})] with $\delta = 0$ and $\gamma_1 = \gamma_2$, ${\rm Re} \ \omega (k)$ is of a folded cosine shape here [Fig. \ref{fig:BHbandstructure1}(a)]. Meanwhile, we notice that the average of the upper and lower real bands is slightly dispersive, and there is hence a tiny SNN coupling component. In this case, the detuning is corrected according to the Rice-Mele Hamiltonian \cite{Rice1982,Longhi2013}, and the mode frequencies read
\begin{equation}
	{\rm Re} \ \omega (k) = \omega_0 - 2 \rho \cos (k L + \phi) \pm 2 \kappa_r \cos \frac{k L}{2}, \label{eq:realfreqBHsystem}
\end{equation}
where $\omega_0$ is the resonance frequency of a single cavity, $\rho \in \mathbb{R}$ and $\phi \in \mathbb{R}$ are the amplitude and additional phase factor of the SNN coupling, respectively. Note that the SNNs of each cavity are a unit cell away, and thus the dispersion by them depends on $kL$. The blue solid curves in Fig. \ref{fig:BHbandstructure1}(a) are the analytic real bands [Eq. (\ref{eq:realfreqBHsystem})] for $\kappa_r = 32.08 \ {\rm GHz}$, $\rho = 0.662 \ {\rm GHz}$ and $\phi = 0.013 \ {\rm rad}$, which fit closely with the simulation result (blue circles). On the other hand, the imaginary bands have narrow but complicated oscillation structures [Fig. \ref{fig:BHbandstructure1}(b)]. Because the imaginary parts of the material indices are all zero, this property should solely be attributed to radiation. The cavity modes here are formed in the thin air-suspended slab. Thus, their small out-of-plane radiation fields result in not only finite on-site loss but also non-local imaginary couplings that give rise to the fast oscillation components in ${\rm Im} \ \omega (k)$.

Complex band structures for cavities with uniform absorption loss, $n_{\rm 1,i} = n_{\rm 2,i} \equiv n_{\rm L,i} > 0$, are systematically investigated. An example with $ n_{\rm L,i} = 0.005$ is plotted in Fig. \ref{fig:BHbandstructure1}(c). Here, simulated ${\rm Re} \ \omega (k)$ (symbols) can be reproduced well by Eq. (\ref{eq:realfreqBHsystem}) with almost the same parameters used for Fig. \ref{fig:BHbandstructure1}(a), as shown again by blue curves (only $\rho$ is slightly changed: $\rho = 0.661 \ {\rm GHz}$). Although ${\rm Im} \ \omega (k)$ for this case still looks wavy, its upper and lower bands become split except for $k = \pi/L$ and are nearly linear around the same point.

As is known in classic laser theory \cite{Sergent1974}, the absorption (carrier excitation by photons) and radiation (coupling with a thermal reservoir) are considered as independent processes. In addition, when the system comprises identical cavities with no loss contrast, it essentially has a single frequency band, where any inter-cavity couplings can only make separable cosinusoidal contributions. Thus, we should be able to extract the contribution of absorption to the simulated ${\rm Im} \ \omega (k)$, by subtracting the imaginary bands induced by radiation [Fig. \ref{fig:BHbandstructure1}(b)] from those involving both the absorption and radiation effects [Fig. \ref{fig:BHbandstructure1}(c)]. The resultant dispersion relative to the average is shown as markers in Fig. \ref{fig:BHbandstructure1}(d). Remarkably, it exhibits a clear two-fold cosinusoidal structure, which is consistent again with Eq. (\ref{eq:bulkeigenvalue1}) for $\delta = 0$ and $\gamma_1 = \gamma_2 = \gamma$, namely
\begin{equation}
	{\rm Im} \ \omega (k) + \gamma = \pm 2 \kappa_i \cos \frac{k L}{2}. \label{eq:imagfreqBHsystem}
\end{equation}
Because of the identity $\int_{0}^{\pi/L} 2 \kappa_i \cos (k L/2) = 4 \kappa_i$, we can determine the absorption-induced imaginary coupling $\kappa_i$ by integrating the discrete data points numerically. Here, we have $\kappa_i = -0.137 \ {\rm GHz}$, and the corresponding analytic curves (solid and red) in Fig. \ref{fig:BHbandstructure1}(d) by Eq. (\ref{eq:imagfreqBHsystem}) indeed agree with the simulation result. The slight discrepancy between them, especially for $k \le 0.2 \pi/L$, is possibly due to a minor contribution of $n_{\rm L,i}$ to $\epsilon_{j,r}$ affecting the mode radiation, or due to some fluctuation of simulation conditions. We also find $\kappa_r > 0$ and $\kappa_i < 0$ from the correspondence between the eigenmode profiles and complex eigenfrequencies for $k = 0$.

\begin{figure}
	\includegraphics[width=7.0cm]{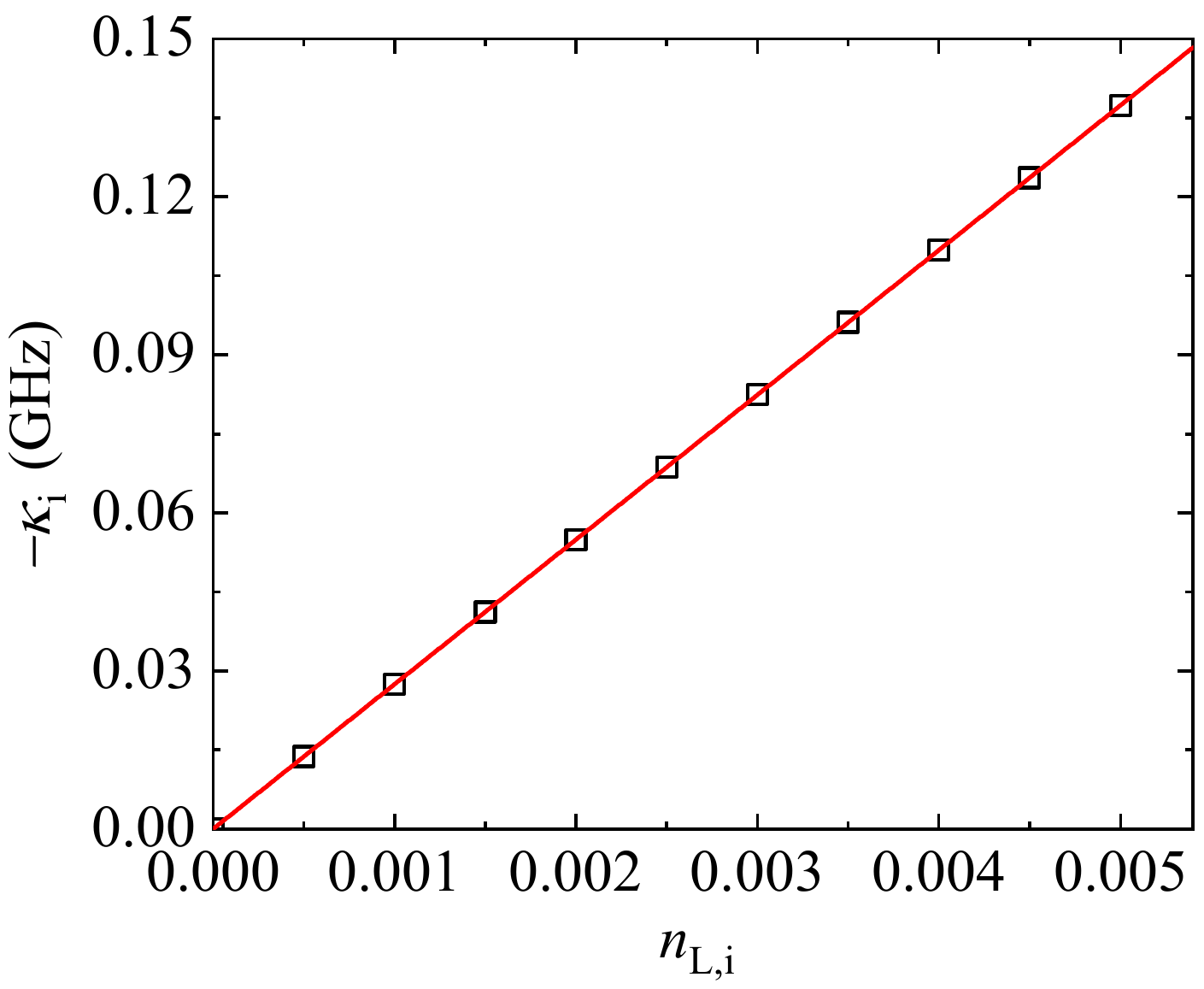}
	\caption{\label{fig:kappaivsni} Simulated imaginary couplings by absorption $\kappa_i < 0$ for different imaginary indices of each heterostructure $n_{\rm L,i}$ (black markers). Least-square fitting (red line) determines their linear dependence as $-\kappa_i = 13.735 \times 2 n_{\rm L,i} \ {\rm GHz}$. As $n_{\rm L,i} \propto \epsilon_{1,i} = \epsilon_{2,i}$, this relation confirms the proportional relation between $\kappa_i$ and $\epsilon_{1,i} + \epsilon_{2,i}$, i.e., Eq. (\ref{eq:imaginarycoupling4}).}
\end{figure}
\begin{figure*}[t!]
	\includegraphics[width=13.0cm]{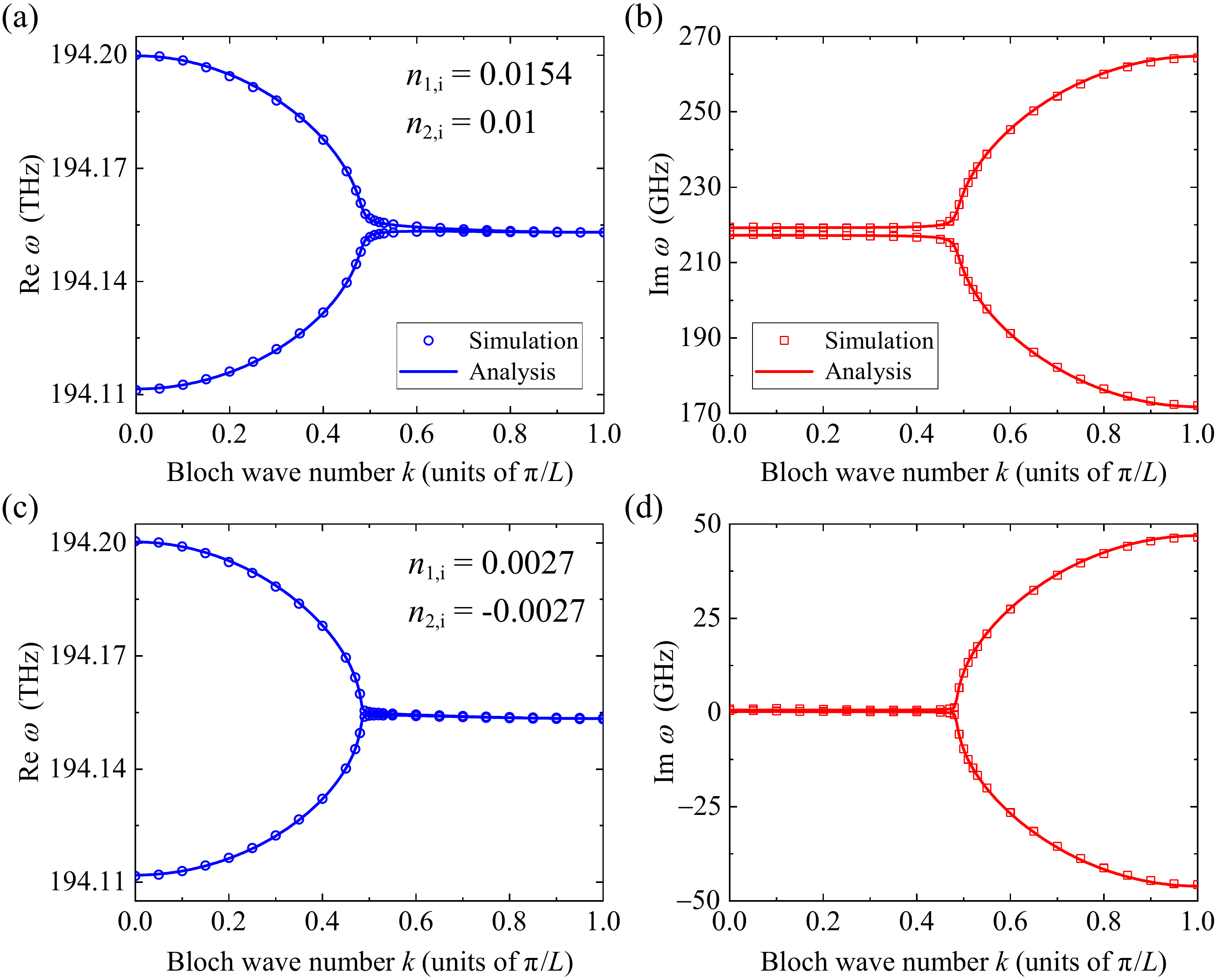}
	\caption{\label{fig:BHbandstructure2} System complex band structures $\omega (k)$ with contrast of gain and loss. (a) ${\rm Re} \ \omega (k)$ and (b) ${\rm Im} \ \omega (k)$ for $(n_{\rm 1,i}, \, n_{\rm 2,i}) = (0.0154, \, 0.01)$. (c) ${\rm Re} \ \omega (k)$ and (d) ${\rm Im} \ \omega (k)$ for $(n_{\rm 1,i}, \, n_{\rm 2,i}) = (0.0027, \, -0.0027)$. Markers: simulation results. Solid curves: analytic results with Eq. (\ref{eq:freqSNN}). $\kappa_r = 32.08 \ {\rm GHz}$, $\phi = 0.04 \ {\rm rad}$, $\rho = 0.66 \ {\rm GHz}$, and $(\gamma_1 - \gamma_2)/2 = -46.5 {\rm GHz}$ corresponding to $n_{\rm 1,i} - n_{\rm 2,i} = 0.0054$. (a) and (b): $\kappa_i = -0.34887 \ {\rm GHz}$, $(\gamma_1 + \gamma_2)/2 = -218.25 \ {\rm GHz}$ and $\omega_0 = 194.1543 \ {\rm THz}$. (c) and (d): $\kappa_i = -0.07 \ {\rm GHz}$, $(\gamma_1 + \gamma_2)/2 = -0.43 \ {\rm GHz}$ and $\omega_0 = 194.1547 \ {\rm THz}$. In (c) and (d), the EP singularity is restored because $\kappa_i$ is suppressed by $n_{\rm 1,i} + n_{\rm 2,i} = 0$. Here, only a small contribution of radiation loss remains.}
\end{figure*}
By repeating the above-mentioned parameter estimation for various $n_{\rm L,i}$, we obtain different $\kappa_i$, as depicted in Fig. \ref{fig:kappaivsni}. The simulated data (black squares) clearly show a relation of proportionality between $\kappa_i$ and $n_{\rm L,i}$, which is confirmed by their regression line (red) giving negligible errors in both the slope and intercept, $-\kappa_i = 13.735 \times 2 n_{\rm L,i} \ {\rm GHz}$. Because $\epsilon_i = {\rm Im} \ (n_r - i n_i)^2 = -2 n_r n_i \propto n_i$ and we use the same material (namely $n_{\rm BH,r}$) for the two heterostructures in the unit, this result strongly supports the notable consequence, Eq. (\ref{eq:imaginarycoupling4}), in our CMT derivation. With Eq. (\ref{eq:ratiocoupling}), we also confirm that $\kappa_r$ and $\kappa_i$ have opposite signs when the system is absorptive, namely $\epsilon_{1,i} + \epsilon_{2,i} < 0$.

We further examine the impact of imaginary couplings on the system under biased and unbiased PT-symmetric configurations. We introduce imaginary index contrast for the cavities, which induces the EP transition in the band structure. When the heterostructures have significant absorption loss and only a small portion of it is compensated, non-negligible $\kappa_i$ will be present and affect the system response. Figure \ref{fig:BHbandstructure2}(a) and (b) display ${\rm Re} \ \omega (k)$ and ${\rm Im} \ \omega (k)$, respectively, for a loss-biased case with $(n_{\rm 1,i}, \, n_{\rm 2,i}) = (0.0154, \, 0.01)$. Here, a previous study \cite{Takata2017} points out that the singularity is most effective when the phase transition occurs near $k = 0.5 \pi/L$ with $|\gamma_1 - \gamma_2|/2 = \sqrt{2} \kappa_r$. $n_{\rm 1,i} - n_{\rm 2,i} = 0.0054$ is close to this condition and corresponds to $(\gamma_1 - \gamma_2)/2 \approx -46.5 {\rm GHz}$. As seen in Fig. \ref{fig:BHbandstructure2}(a) and (b), the EP degeneracy is lifted, and the divergence of the complex differential frequency ${\rm d}\omega/{\rm d}k$ around the coalescence of ${\rm Re} \ \omega (k)$ is significantly suppressed in this lossy system. Moreover, ${\rm Im} \ \Delta \omega (k)$ in the exact phase ($k \lesssim 0.5 \pi/L$) are split. These features indicate the existence of finite $\kappa_i$.

Remarkably, our non-Hermitian CMT keeps consistency with the simulation result in Fig. \ref{fig:BHbandstructure2}(a) and (b). The theoretical eigenfrequencies for the system with $\delta = 0$ and a finite SNN coupling are given by
\begin{align}
	\omega (k) = \omega_0 &- 2 \rho \cos (k L + \phi) - i \frac{\gamma_1 + \gamma_2}{2} \nonumber \\
	&\pm \sqrt{\bigg[ 2 \left(\kappa_{r} + i \kappa_{i}\right) \cos \frac{k L}{2} \bigg] ^2 - \bigg(\frac{\gamma_1 - \gamma_2}{2} \bigg)^2 }. \label{eq:freqSNN}
\end{align}
Here, we already have the information of $\omega_0$, $\kappa_r$, $\rho$ and $\phi$ from Fig. \ref{fig:BHbandstructure1}, and the average of ${\rm Im} \ \Delta \omega (k)$ in Fig. \ref{fig:BHbandstructure2}(b) gives $(\gamma_1 + \gamma_2)/2 = -218.25 \ {\rm GHz}$. Thus, the only unknown parameter in Eq. (\ref{eq:freqSNN}) is $\kappa_i$. Now, we \textit{predict} its value by the extrapolation of Fig. \ref{fig:kappaivsni} for the general case with $n_{\rm 1,i} \ne n_{\rm 2,i}$. Equation (\ref{eq:imaginarycoupling4}) and the slope of Fig. \ref{fig:kappaivsni} suggest $\kappa_i = -13.735 \times (n_{\rm 1,i} + n_{\rm 2,i}) = -0.34887 \ {\rm GHz}$ for $(n_{\rm 1,i}, \, n_{\rm 2,i}) = (0.0154, \, 0.01)$. We draw analytic curves of $\omega (k)$ in Fig. \ref{fig:BHbandstructure2}(a) and (b) with the collected parameters including $\kappa_i$. They agree well with the simulated eigenfrequencies shown as symbols. This indicates that our CMT is valid for a system with a broad range of gain and loss based on the cavity media.

If $\kappa_i$ is proportional to $n_{\rm 1,i} + n_{\rm 2,i}$, we should be able to cancel it by setting $n_{\rm 2,i} = - n_{\rm 1,i}$. Fig. \ref{fig:BHbandstructure2}(c) and (d) depict $\omega (k)$ for $(n_{\rm 1,i}, \, n_{\rm 2,i}) = (0.0027, \, -0.0027)$ and show that this is indeed the case. Here, the abrupt coalescence of the complex bands is restored, as compared to Fig. \ref{fig:BHbandstructure2}(a) and (b). Because we keep the value of $n_{\rm 1,i} - n_{\rm 2,i}$ and hence $(\gamma_1 - \gamma_2)/2$, it is only the change in $\kappa_i$ that affects the radical term in Eq. (\ref{eq:freqSNN}), namely the EP transition. Thus, the singular spectral behavior directly reflects the suppression of the imaginary coupling by the balanced gain and loss. We also find that theoretical curves with $(\gamma_1 + \gamma_2)/2 = -0.43 \ {\rm GHz}$ and reduced $\kappa_i = -0.07 \ {\rm GHz}$ successfully reproduce the simulation result (markers) including its fine structure around $k = 0.5 \pi/L$. Here, the former parameter is actually taken from the imaginary band edge in Fig. \ref{fig:BHbandstructure1}(b), and the latter is within the order of the loss splitting in the same plot. Thus, we regard the remaining $\kappa_i$ as the NN approximation of the radiation effects. 

Overall, our theoretical model and method compose a powerful tool to predict the behavior of photonic coupled-mode systems with amplification and absorption. The imaginary coupling by the gain media is deterministic, and it can be canceled with balanced gain and loss. 

\section{\label{sec:RBEP} Radiation-induced exceptional points in photonic crystal nanocavities}
\begin{figure}[t!]
	\includegraphics[width=8.6cm]{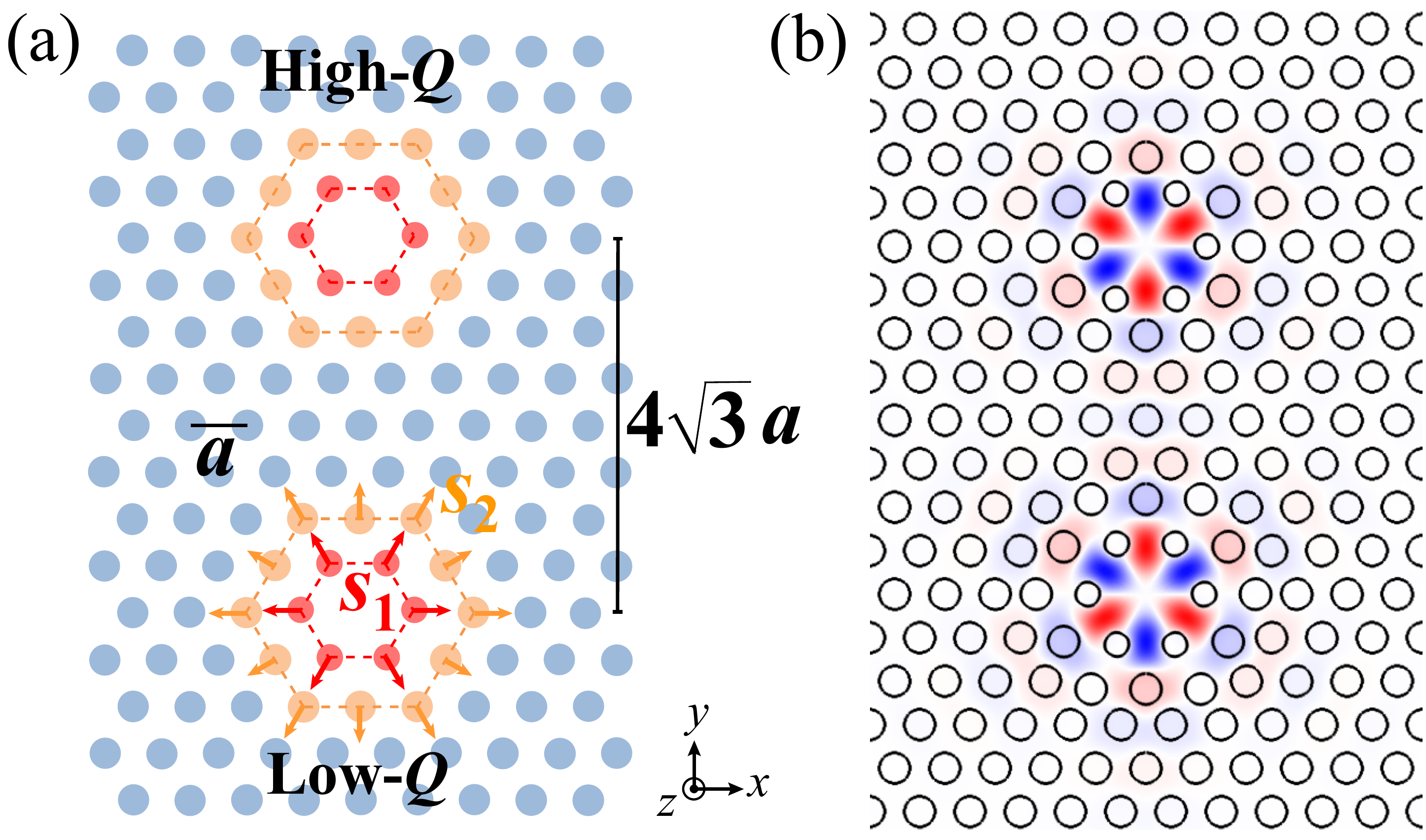}
	\caption{\label{fig:RadiationH1System} (a) Two coupled Si H1 photonic crystal cavities with contrast of radiation loss. Cavity 1 (upper) has a solitary theoretical $Q$ factor over $10^8$. The loss rate for cavity 2 (lower) is varied by displacing air holes away from the point defect. $s_1$ and $s_2$: hole shifts for the innermost and second innermost layers from the lattice-matched position, respectively. (b) One of the considered hexapole supermodes.}
\end{figure}

In this section, we show a way to design a clean radiation-based EP and the mechanism of its formation in a realistic on-chip device. We simulate and analyze the two-cavity system illustrated in Fig. \ref{fig:RadiationH1System}(a), which only contains the NN coupling. The system comprises an air-suspended Si photonic crystal slab and two point-defect nanocavities \cite{Notomi2004}. Here, the upper and lower cavities, cavity 1 and 2, respectively, are separated by a distance of $d = 4 \sqrt{3}a$, where $a = 426 \ {\rm nm}$ is the hole period. The slab thickness is 250 nm, and the refractive index of Si is set as 3.47. The system has 9 and 11 barrier layers on each side of the photonic molecule in $x$ and $y$ directions, respectively. The air holes of the photonic crystal have radius $R_0 = 131 \ {\rm nm}$, while those closest the point defects are of smaller size, $R_1 = 102 \ {\rm nm}$.

For controlling the mode frequencies and radiation loss, both cavities involve spatial shifts of their innermost and second innermost shells of air holes directed away from their centers [red and orange ones in Fig. \ref{fig:RadiationH1System}(a)], with the regular hexagonal hole alignment kept. For cavity 1, the first and second shells are constantly broadened in their half diagonals by 89.5 nm and 20.5 nm from the lattice-matched position, respectively. As a result, we find a hexapole mode with a ultrahigh theoretical $Q$ factor of $Q = 1.4 \times 10^8$; details will be studied numerically and experimentally elsewhere \cite{TakataInprep}. We also adjust the shifts of the inner and outer layers for cavity 2, denoted as $s_1$ and $s_2$ respectively, so that it has much larger radiation loss but keeps its frequency close to that of cavity 1. The $z$ component of the magnetic fields for one of the coupled modes is depicted in Fig. \ref{fig:RadiationH1System}(b).

\begin{figure}[t!]
	\includegraphics[width=8.6cm]{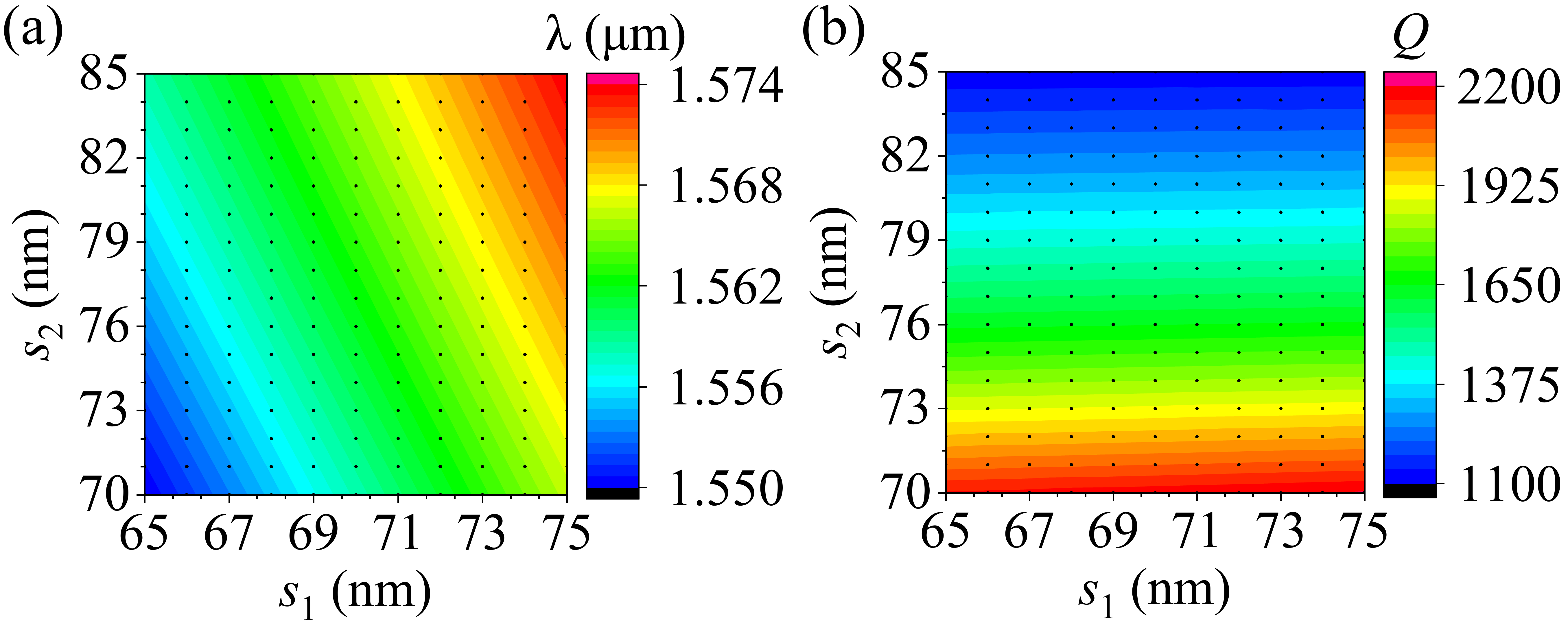}
	\caption{\label{fig:LossyCavityFQ} (a) Wavelength $\lambda$ and (b) $Q$ factor of the single hexapole cavity mode depending on $s_1$ and $s_2$. Because of the distinct gradients of $\lambda$ and $Q$, we can change them independently with $s_1$ and $s_2$.}
\end{figure}

Fig. \ref{fig:LossyCavityFQ}(a) and (b) show the dependence of the wavelength $\lambda$ and $Q$ factor on $s_1$ and $s_2$ for the cavity 2's hexapole mode in the solitary environment. Here, $(\lambda, Q)$ change monotonically with $(s_1, s_2)$, and thus they have one-to-one correspondence in our entire simulation result. Fig. \ref{fig:LossyCavityFQ}(a) has diagonal iso-wavelength lines, since both parameters affect the cavity shape and hence $\lambda$. In contrast, $Q$ is dominated by $s_2$ in this low-$Q$ regime, as seen in Fig. \ref{fig:LossyCavityFQ}(b). We have found that it can be as small as $Q \approx 500$ before $s_2$ becomes large enough for the second shell to merge with other air holes. The result shows that we can achieve wide-range and independent control of the resonance detuning $\delta$ and loss contrast $l = (\Gamma_1 - \Gamma_2)/2$ ($-\Gamma_i$: on-site radiation loss of cavity $i$), by varying $s_1$ and $s_2$.

The difference in the system eigenfrequencies, $\Omega = \Delta \omega_{+} - \Delta \omega_{-}$, helps us demonstrate an ideal EP transition in our simulation. With Eq. (\ref{eq:twocaveigenvalue2}), we see that it only includes the radical term, namely
\begin{equation}
	\Omega = 2 \sqrt{\kappa^2_{r} - K^2_{i} - l^2 + \delta^2 + 2i (\kappa_{r} K_{i} - l \delta)}, \label{eq:evdifference}
\end{equation}
where $\kappa_r$ and $K_{i}$ are the real and imaginary couplings defined in Eq. (\ref{eq:realcoupling3}) and Eq. (\ref{eq:imcouplingradiation}), respectively. Eq. (\ref{eq:evdifference}) hence gives the information of the corresponding unbiased system discussed in Sec. \ref{sec:CMTRad}. Remarkably, simulated eigenvalues with a purely real or imaginary $\Omega$ satisfy Eq. (\ref{eq:generalPTcondition}), thereby resulting in the general PT symmetry. We seek for such data points so that we can find an exact EP on this continuous parametric curve mapped onto the $(s_1, s_2)$ plane.

When we modify Eq. (\ref{eq:generalPTcondition}) as
\begin{equation}
	K_{i} = \frac{l \delta}{\kappa_{r}}, \label{eq:Ki}
\end{equation}
we can also solve for $\kappa_r$ by substituting Eq. (\ref{eq:Ki}) back into Eq. (\ref{eq:evdifference}),
\begin{equation}
	\kappa_{r} = \pm \frac{1}{\sqrt{2}} \sqrt{\frac{\Omega^2}{4} - l^2 - \delta^2 + \sqrt{\left(\frac{\Omega^2}{4} - l^2 - \delta^2 \right)^2 + 4l^2 \delta^2} }. \label{eq:kappar}
\end{equation}
This means that the data with $\Omega \in \mathbb{R}$ or $\Omega \in i\mathbb{R}$ impose an additional constraint and hence enable us to estimate $\kappa_{r}$ and $K_{i}$ with Eqs. (\ref{eq:Ki}) and (\ref{eq:kappar}), in combination with $(l, \delta)$ obtained in an additional simulation for the corresponding single-cavity conditions.
\begin{figure*}
	\includegraphics[width=14.0cm]{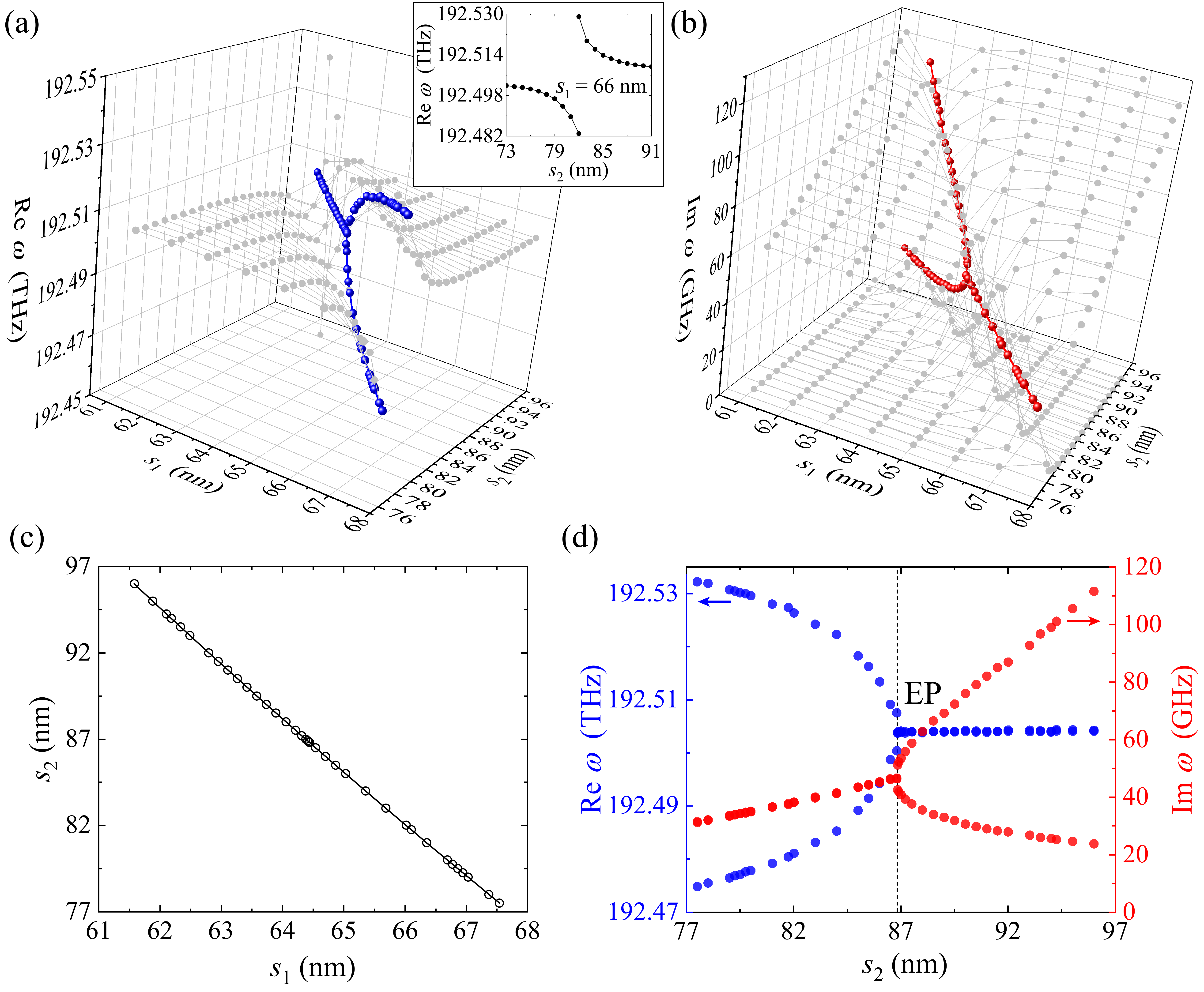}
	\caption{\label{fig:H1CavitiesEP} EP phase transition induced only by the radiation loss in the simulated system. (a) ${\rm Re} \ \omega_{\pm}$ and (b) ${\rm Im} \ \omega_{\pm}$ of the eigenfrequencies $\omega_{\pm}$ around the EP. Colored symbols: solutions with ${\rm Im} \ \omega_{+} \approx {\rm Im} \ \omega_{-}$ or ${\rm Re} \ \omega_{+} \approx {\rm Re} \ \omega_{-}$, which means $\Omega \in \mathbb{R}$ or $\Omega \in i\mathbb{R}$ within an error of 0.3 GHz. Inset of (a): ${\rm Re} \ \omega_{\pm}$ on a slice of the parameter space with $s_1 = 66 \ {\rm nm}$, which exhibits an avoided crossing. (c) Parameters $(s_1, s_2)$ yielding the colored plots in (a) and (b). (d) Complex eigenfrequencies on the trajectory shown in (c) as a function of $s_2$. An almost ideal EP is formed, despite that the system is affected by the imaginary coupling $K_i$ arising from radiation.}
\end{figure*}

Simulation of our two-cavity device reveals an ideal radiation loss-based EP. Fig. \ref{fig:H1CavitiesEP}(a) and (b) depicts the real and imaginary parts of the pairwise eigenfrequencies $\omega =\omega_0 + \Delta \omega_{\pm}$ on the $(s_1, s_2)$ plane. Here, we highlight a series of solutions that have negligible ${\rm Re} \ \Omega$ or ${\rm Im} \ \Omega$ ($< 0.3 {\rm GHz}$) in blue and red in (a) and (b), respectively. When the solitary resonance frequency of cavity 2's hexapole mode is close to that of cavity 1, the system exhibits a strong coupling. We can hence see the resultant avoided crossing of the two frequency branches for relatively large $s_1$ and small $s_2$ [inset of Fig. \ref{fig:H1CavitiesEP}(a)]. Sweeping $s_1$ and $s_2$ with fine resolutions in this region enables us to find the solutions with ${\rm Im} \ \Omega \approx 0 \ (\Omega \in \mathbb{R})$, i.e., evenly distributed coupled modes with the same net loss. By carefully tracing such states along with an iso-wavelength (frequency) line for larger $s_2$ (and loss $|\Gamma_2|$), the eigenvalues with real splittings $\Omega \in \mathbb{R}$ coalesce and turn into those with imaginary ones $\Omega \in i\mathbb{R} \ ({\rm Re} \ \Omega \approx 0)$. This process is shown as the colored data, which thereby demonstrate the general PT phase transition with Eq. (\ref{eq:generalPTcondition}). The flat spectrum away from the strong coupling domain in Fig. \ref{fig:H1CavitiesEP}(a) comes from the static cavity 1 and corresponds to the lower ${\rm Im} \ \omega$ in Fig. \ref{fig:H1CavitiesEP}(b). The dispersive solution based on cavity 2 is distributed outside the plot range of Fig. \ref{fig:H1CavitiesEP}(a) and has the higher ${\rm Im} \ \omega$.

Fig. \ref{fig:H1CavitiesEP}(c) shows the actual trajectory of the parameters that give $\Omega \in \mathbb{R}$ or $\Omega \in i\mathbb{R}$ in the simulation. Here, $s_1$ is adjusted in 0.001 nm units for acquiring the data points shown by markers. The solid curve is the least-square quadratic interpolation of the simulation result, $s_1 = 1.182685 \times 10^{-3} s_2^2 - 5.276275 \times 10^{-1} s_2 + 1.013309 \times 10^{2}$. $s_1$ and $s_2$ are in nanometers, and its coefficient of determination $R^2$ satisfies $1 - R^2 \approx 10 ^{-8}$.

Holding this relation of $(s_1, s_2)$, the complex eigenfrequencies under the EP transition are redrawn as a function of $s_2$ in Fig. \ref{fig:H1CavitiesEP}(d). By the change in $s_2$ and $\Gamma_2$, a nearly strict EP is formed at $(s_1, s_2) = (64.431 \ {\rm nm}, 86.84 \ {\rm nm})$. Here, ${\rm Re} \ \Delta \omega_{\pm}$ reach the singular coalescence without notable residual splitting, and ${\rm Im} \ \Delta \omega_{\pm}$ bifurcate sharply from the coincident branches. We actually need extra care to achieve such clean properties of the EP even in simulation, when realistic structures are considered \cite{Takata2017,Nguyen2016,Yao2019,Benisty2012}. This is mostly due to persisting imaginary couplings, and the question we answer here is how the EP is restored in our system whose only non-Hermitian factor is radiation.

\begin{figure*}
	\includegraphics[width=17.6cm]{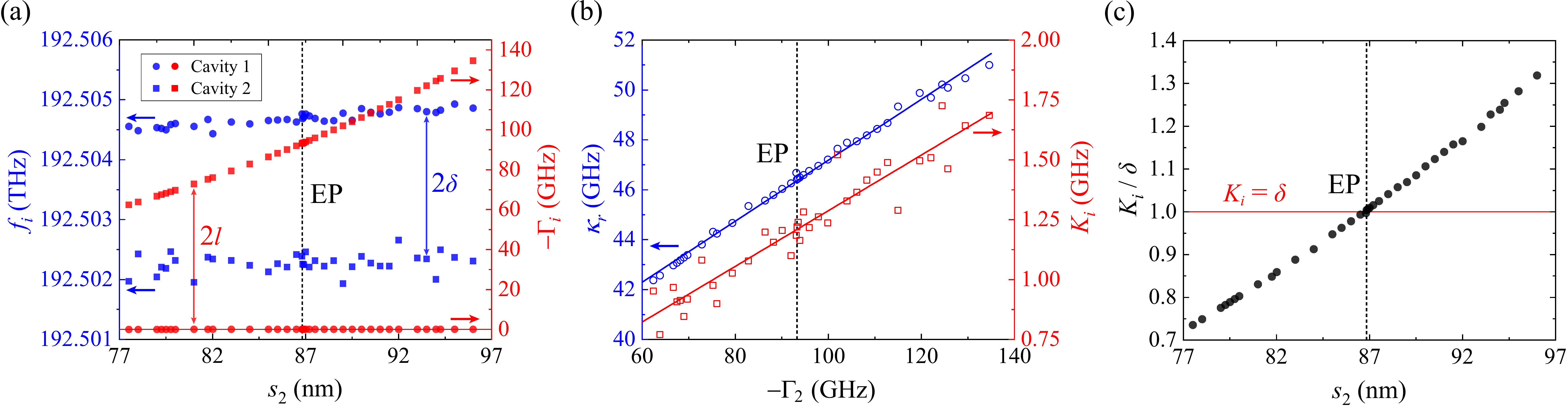}
	\caption{\label{fig:KIdlCompensation} (a) Simulated frequency $f_i$ and loss rate $-\Gamma_i$ for the hexapole mode of each isolated cavity ($i = 1, 2$), revealing steady resonance detuning $2\delta$ in the EP transition [Fig. \ref{fig:H1CavitiesEP}(d)]. Here, $(s_1, s_2)$ of cavity 2 follow the plot in Fig. \ref{fig:H1CavitiesEP}(c). Each result includes the correction by the shift of the second innermost air hole layer for the \textit{other} cavity. Cavity 1 holds $Q$ factors over $10^6$. (b) Complex cavity coupling $\kappa_r + i K_i$ estimated with (a) as a function of $\Gamma_2$. The radiation loss also promotes the in-plane spatial mode broadening and hence enhances both $\kappa_r$ and $K_i$. (c) The ratio between $K_i$ and $\delta$ in the EP transition. It increases with $s_2$ and reaches unity at the EP, namely $K_i = \delta$, confirming the compensation of the imaginary coupling by the detuning there.}
\end{figure*}

To analyze the spectrum of the coupled eigenmodes, we perform another series of simulations for the frequency $f_i = \omega_0 \pm \delta$ and loss rate $-\Gamma_i$ of each cavity mode ($i = 1, 2$). The result for the parameter points of Fig. \ref{fig:H1CavitiesEP}(c) is shown in Fig. \ref{fig:KIdlCompensation}(a) as a function of $s_2$. Here, we notice that the cavities in the coupled system are so proximate that structural modulation for one cavity affects the resonance of the other. Thus, we include the shift of the second innermost hole shell by $s_2$ for cavity 2 in simulating cavity 1, and vice versa with the fixed layer displacement of 20.5 nm for cavity 1, to obtain better accuracy; the condition is hence different from Fig. \ref{fig:LossyCavityFQ}. Remarkably, the mode frequency $f_1$ of cavity 1 is consistently higher than that of cavity 2 ($f_2$), revealing finite cavity detuning $2\delta = f_1 - f_2 \approx 2.5 \ {\rm GHz}$ over the entire EP transition. 
The on-site loss $-\Gamma_2$ of cavity 2 is significantly varied via $s_2$ in the range between 60 and 140 GHz. On the other hand, cavity 1 holds $Q > 10^6$, despite that $s_2 > 77 \ {\rm nm}$ means highly lattice-mismatched scattering defects located nearby.

With Fig. \ref{fig:H1CavitiesEP}(d), Fig. \ref{fig:KIdlCompensation}(a), and Eqs. (\ref{eq:Ki}) and (\ref{eq:kappar}) at hand, we can now calculate the dependence of $\kappa_r + i K_i$ on $-\Gamma_2 > 0$, as shown in Fig. \ref{fig:KIdlCompensation}(b). Because $\Gamma_1$ is negligible and $\delta$ varies little, the trend in the complex coupling is attributed to the major loss factor $-\Gamma_2$. $\kappa_r$ is positive in this case, since the anti-symmetric eigenmode $(-1, 1)^{\rm T}$ [Fig. \ref{fig:RadiationH1System}(b)] has the higher frequency corresponding to $\omega_0 + \Delta \omega_{+}$. We also find $K_i > 0$ due to $l = (\Gamma_1 - \Gamma_2)/2 > 0$ in Eq. (\ref{eq:Ki}). $\kappa_r$ as well as $K_i$ has a positive correlation with $-\Gamma_2$, because enhancing radiation also involves in-plane spatial mode broadening. As such, this result arises not from material properties but from the change in the cavity mode profiles $\{\bm{\Phi}_{1}, \bm{\Phi}_{2}\}$. The point closest to the EP has $l = 46.61 \ {\rm GHz}$ and $\kappa_r = 46.41 \ {\rm GHz}$, which indicate $\kappa_r = l$ in Eq. (\ref{eq:EPCondition}) within the error of 0.3 GHz. Solid curves provide best-fit regression lines for the data: $\kappa_r = -0.1222 (1.425 \times 10^{-3}) \Gamma_2 + 34.96 (0.1373) \ {\rm GHz}$, and $K_i = -1.161 \times 10^{-2} (6.254 \times 10^{-4}) \Gamma_2 + 0.1261 (6.025 \times 10^{-2}) \ {\rm GHz}$, which indicate clear correlation among $(\kappa_r, K_i, \Gamma_2)$ in the broad parameter range of the plot. Nonetheless, we do not intend to identify the global dependence of the coupling terms on the parameters with the linear regression. Because the change in $|\Gamma_2|/2$ is larger than that in $\kappa_r$, we can find the EP as long as the process is continuous. The values of $K_i$ are small and thus fluctuated by subtle parameter and meshing conditions in the finite-element simulation.

Because the variation in the detuning is actually correlated with that in the imaginary coupling, we can obtain a consistent transition of their ratio $K_i/\delta$ in the process, as shown in Fig. \ref{fig:KIdlCompensation}(c). Here, $K_i/\delta$ is dominated by $K_i$ and increases with $-\Gamma_2$ and hence $s_2$. Remarkably, the plot crosses $K_i/\delta = 1$ precisely at the EP, namely
\begin{equation}
	K_i = \delta, \label{eq:Kidcompensation}
\end{equation}
clarifying the cancellation of the effect of imaginary coupling by the balanced detuning. Equation (\ref{eq:Kidcompensation}) together with $\kappa_r = l$ is indeed the condition for the system to reach an EP, Eq. (\ref{eq:EPCondition}). This means that our simulation result is fully explained within the framework of our CMT.

\section{\label{sec:conclusion} Discussion and Conclusion}
The non-Hermitian CMT derived here will be applicable for many systems of evanescently coupled dielectric resonators, such as stripe lasers \cite{Yao2019,Zhu2018} and VCSELs \cite{Gao2017,Gao2019}. The imaginary couplings in whispering-gallery-mode cavities, such as ring and disk resonators \cite{Peng2014,Chang2014,Feng2014sin,Hodaei2014}, will also be obtainable by taking into consideration pairwise circulation of the modes. 

The validity of the CMT is based on the condition that the basis modes are not significantly disrupted. Quantitatively, it is guaranteed by the fact that the evanescent coupling ($\kappa_r \lesssim 50 \ {\rm GHz}$) and gain and loss ($|\gamma_l|, \ |\Gamma_i| \lesssim 260 \ {\rm GHz}$) are much smaller than the cavity-mode frequencies ($\sim 200 \ {\rm THz}$) in our simulation. Here, by modeling the radiation with the huge but closed system including the air and virtual absorber, our CMT has been explicitly shown to cover the radiation loss and radiation-induced imaginary couplings. Our result indicates that it can predict the behavior of dielectric cavities with amplification, absorption, and radiation, which exhibit net $Q$ factors of several hundreds and larger, within the scope of the Maxwell equations.

On the other hand, for metallic nanoresonators and nanoantennas based on surface-plasmon polaritons, their modal properties are determined intrinsically by the striking kinetic loss of carriers that demands $Q \approx 10$. In this case, lossless basis can no longer be prepared, and thus coupled modes have to be expanded by states with complex frequencies, which are termed quasinormal modes (QNMs) \cite{Lalanne2018}. A prominent signature of QNMs is the divergence of far fields, and such QNM systems involve the modification of the analytic formulae for the mode volume and Purcell factor \cite{Sauvan2013}. This QNM formalism is also essential for very leaky optical resonators \cite{Muljarov2016}.

We have not detected visible features peculiar to QNMs in our simulation. In addition, the first-order CMT reproduced the experimental spontaneous emission spectrum of coupled high-$Q$ photonic crystal lasers operating near a weakly loss-biased EP \cite{Takata2021}. However, we notice that a rigorous quantum-mechanical treatment of spontaneous emission, which is beyond the conventional Fermi's golden rule based solely on the photonic local density of states, is necessary, especially when the system has gain \cite{Franke2021}. Another exception to our CMT is systems of electromagnetic resonators, where electric and magnetic fields are coupled via surface current densities \cite{Park2020, Elnaggar2015}.

We have clarified that biased PT-symmetric cavities with large total loss can suffer from a pronounced effect from the imaginary couplings $\{\kappa_i + K_i\}$. By compensating their loss and applying gain to half of them with pumping, the permittivity-induced component $\{\kappa_i\}$ can be suppressed drastically. In contrast, the radiation-based factor $\{K_i\}$ might not be necessarily canceled even for lasers, since it depends crucially on the cavity mode profiles and their arrangement. In particular, the behavior of imaginary couplings in two-dimensional arrays, including possible non-local factors, should be investigated in detail.

We can find a rigorous radiation-based EP by adjusting both the cavity detuning $\delta$ and loss contrast $l$. In our simulation, the spatial shifts of the air holes were finely controlled. Technically available resolutions of the hole position are about 0.1 nm; thus, experimental demonstration of the device may result in a spectrum that is slightly off from the exact EP. However, resonance linewidths of loss-biased coupled modes near the EP are dominated by the lossy cavity and thus broad enough to cover such discrepancy. Representative EP responses, such as spectral coalescence and unidirectional reflectivity, are hence expected to be observed.

In conclusion, we established the coupled-mode theory for optical cavities with amplification, absorption, and radiation. We analytically determined the imaginary couplings between cavities with different imaginary permittivities. We also presented, to our knowledge, the first explicit model and physical implications of the radiation-induced imaginary coupling terms in the coupled-mode formalism. They have equal forward and backward components for the basis bound modes, representing non-Hermiticity of the system.

Regardless of their origins, the imaginary couplings can lift the EP degeneracy. Thus, their impact should be measured in practical systems, and it is necessary to find out how to counteract them and restore the EP. We provided a scheme to precisely estimate the permittivity-induced imaginary coupling and confirmed its properties with a simulation of a periodic array of buried-heterostructure nanocavities. Because this factor is proportional to the sum of the imaginary parts of dielectric constants for adjacent cavity media, it can be suppressed by their balanced gain and loss. We also identified the radiation-induced imaginary coupling as the contribution of the cavity modes' interference to their net radiation loss. In systems of two resonators with contrast of radiation loss, the EPs should remain protected by the general PT symmetry. Our simulation actually revealed the ideal radiation-based EP of the two H1 Si photonic crystal cavities. Here, we confirmed that not only the real coupling and loss contrast but also the imaginary coupling and cavity detuning were balanced at the EP. 

There have been several theoretical approaches for clarifying the disruption and restoration of the EP in practical systems. In an early study, the imaginary coupling was introduced as a phenomenological term \cite{Benisty2012}. A perturbation-analysis formalism for two waveguides \cite{Nguyen2016} presented permittivity-induced complex couplings that lifted the EP, but they seemed asymmetric in terms of the guided-mode basis. Another CMT derived imaginary couplings in systems with uniform absorption loss \cite{Golshani2014}. Our CMT extends Ref. \onlinecite{Golshani2014} and covers all major gain and loss mechanisms in lasers. It enables us to determine the imaginary couplings as well as other parameters in simulations of practical systems, and the resultant analysis will give consistent complex eigenfrequencies that let us identify the EP protected by the generalized PT symmetry. This work hence establishes dependable design principles for photonic devices with EPs. In addition, the imaginary coupling has potential as an additional degree of freedom, which would elevate non-Hermitian state control and nonlinear effects. 

\begin{acknowledgments}
	We thank Yasuhiro Hatsugai, Tsuneya Yoshida, Yuto Moritake, and Taiki Yoda for fruitful discussions. We acknowledge the research placement program of University of Bath for supporting this project. This work was supported by JSPS KAKENHI Grant Number 20H05641.
\end{acknowledgments}




\providecommand{\noopsort}[1]{}\providecommand{\singleletter}[1]{#1}%

\end{document}